# Posterior-based proposals for speeding up Markov chain Monte Carlo

C. M. Pooley[1,2], S. C. Bishop[3], A. Doeschl-Wilson[1] and G. Marion[2]

1 The Roslin Institute, The University of Edinburgh, Midlothian, EH25 9RG, UK.

2 Biomathematics and Statistics Scotland, James Clerk Maxwell Building, The King's Buildings, Peter Guthrie Tait Road, Edinburgh, EH9 3FD, UK.

3 Deceased.

Corresponding author: Dr C. M. Pooley

Address: The Roslin Institute, The University of Edinburgh, Midlothian, EH25 9RG, UK

Email: christopher.pooley@roslin.ed.ac.uk

Tel: +44 (0)131 651 9100

ORCHID: 0000-0002-8779-4477

# Abstract

Markov chain Monte Carlo (MCMC) is widely used for Bayesian inference in models of complex systems. Performance, however, is often unsatisfactory in models with many latent variables due to so-called poor mixing, necessitating development of application specific implementations. This paper introduces "posterior-based proposals" (PBPs), a new type of MCMC update applicable to a huge class of statistical models (whose conditional dependence structures are represented by directed acyclic graphs). PBPs generates large joint updates in parameter and latent variable space, whilst retaining good acceptance rates (typically 33%). Evaluation against other approaches (from standard Gibbs / random walk updates to state-of-the-art Hamiltonian and particle MCMC methods) was carried out for widely varying model types: an individual-based model for disease diagnostic test data, a financial stochastic volatility model, a mixed model used in statistical genetics and a population model used in ecology. Whilst different methods worked better or worse in different scenarios, PBPs were found to be either near to the fastest or significantly faster than the next best approach (by up to a factor of 10). PBPs therefore represent an additional general purpose technique that can be usefully applied in a wide variety of contexts.

# 1 Introduction

Markov chain Monte Carlo (MCMC) techniques allow correlated samples to be drawn from essentially any probability distribution by iteratively generating successive values of a carefully constructed Markov chain. This flexibility has led MCMC to become the method of choice for inferring model parameters under Bayesian inference [1]. However, for high dimensional systems (*e.g.* where inference is over many tens, hundreds or even thousands of variables), MCMC often suffers from a problem known as "poor mixing". This manifests itself as a high degree of correlation between consecutive samples along the Markov chain, so requiring a very large number of iterations to adequately explore the posterior [1]. This limitation is of practical importance, because it restricts the possible models to which MCMC can realistically be applied. The focus of this paper is to introduce and explore a new approach that helps alleviate these mixing problems, thus reducing the computational time necessary to generate accurate inference. This approach has practical advantages over existing methodologies that aim to address the same problem [2-4].

In Bayesian inference the posterior distribution $\pi(\theta,\xi|y)$ represents the state of knowledge concerning the parameters $\theta$ and latent variables $\xi$ of a given stochastic model taking into account data $y$. Using Bayes' theorem this posterior distribution can be expressed as

$$\pi(\theta,\xi|y) = \frac{\pi(y|\xi,\theta)\pi(\xi|\theta)\pi(\theta)}{\pi(y)}, \tag{1}$$

where $\pi(y|\xi,\theta)$ is here referred to as the observation model, $\pi(\xi|\theta)$ is the latent process (*i.e.* the part of the model which is not directly observed but helps explains the observations) likelihood, $\pi(\theta)$ is the prior distribution (representing the state of knowledge prior to data $y$ being considered), and $\pi(y)$ is a constant factor known as the model evidence [5].

An MCMC implementation of Bayesian inference aims to produce samples from the posterior, *i.e.* a list of parameter values $\theta^i$ and latent variables $\xi^i$ distributed in accordance with Eq.(1). This is achieved by sequentially proposing some change to the current state $\theta^i$ and/or $\xi^i$ to generate $\theta^p$ and $\xi^p$, and accepting or rejecting this change with a Metropolis-Hastings (MH) probability to create the next member on the list, *i.e.* $\theta^{i+1}$ and $\xi^{i+1}$ [6]. Note, in some instances it is possible to probabilistically generate samples directly, *e.g.* via Gibbs sampling [7] or slice sampling [8], without the need for an accept/reject step[1]. The term Monte Carlo refers to the probabilistic nature of these updates that form a Markov chain, *i.e.* step $i$+1 in the chain only depends on the state at the previous step.

A typical approach is to sequentially update each parameter and latent variable separately. Figure 1(a) illustrates the problem of poor mixing resulting from implementing this "standard" MCMC when there is a strong dependency between latent variables and model parameters. The dark shaded area represents high posterior probability[2] as a function of $\theta$ and $\xi$. Consider first fixing $\theta$ and making changes to $\xi$, as shown by the dashed line in Fig. 1(a). Because MCMC samples are probabilistically constrained to lie in the shaded region, the chain will make limited progress even for a large number of MH updates. Similarly, changes to $\theta$ whilst fixing $\xi$ will be restricted to move along horizontal lines, which are again limited in scope. A typical output from an MCMC algorithm which independently updates parameters and latent variable is illustrated in Fig. 1(b), which shows the trace plot for one of the variables in θ. Thousands of MH updates are potentially needed to generate just one uncorrelated effective sample from the posterior.

---

[1] In these cases the MH acceptance probability is exactly one.
[2] Note, this diagram is purely schematic, as $\theta$ and $\xi$ are usually multi-dimensional quantities.

The way out of this sorry state of affairs is shown in Fig. 1(c). Here the proposals are performed *jointly* in parameter and latent variable space (as indicated by the pink shaded region) and share a similar correlated structure to the posterior itself. In this case, proposals can jump much further without the posterior probability becoming negligibly small. Consequently, as illustrated in Fig. 1(d), successive samples are less correlated and fewer are needed to be representative of the posterior.

Various techniques to perform joint updates have been proposed in the literature. For example Particle MCMC (PMCMC) [2] samples a new set of parameters $\theta^p$ relative to $\theta^i$ and then sets about generating $\xi^p$. This is achieved by directly sampling from the model $\pi(\xi|\theta^p)$ multiple times, with each instance referred to as a "particle". The final result is built up in a series of stages that sequentially take into account a larger fraction of the data $y$. At the end of each stage those particles which agree well with the sequentially introduced observations are duplicated at the expense of those which don't agree so well (so-called particle filtering). Whilst this method exhibits very good mixing, computational speed is often compromised because the number of particles needed to generate a reasonable acceptance probability can be very large[3] [9].

Approximate Bayesian Computation (ABC) [3] also samples directly from the model, but rather than fitting the data to the samples through a full observation model, as in Eq.(1), here the fit with the data is characterised by a much simpler distance measure $\chi$. Often $\chi$ is specifically chosen such that a substantial proportion of simulated samples contribute to the final result. Such an approach, however, comes at a significant cost, because ABC generates only approximate, rather than exact, draws from the posterior [10]. Furthermore, doubts have been cast on ABC's ability to accurately perform model selection [10, 11].

In Hamiltonian MCMC (HMCMC) [3, 12] $\theta^i$ and $\xi^i$ are dynamically changed through a series of small intermediary steps (which take into account local gradients in the log of the posterior probability) to reach $\theta^p$ and $\xi^p$. This final state is then accepted or rejected with an overall MH probability. This, again, produces good mixing (due to the fact that $\theta^i,\xi^i$ and $\theta^p,\xi^p$ can be widely separated), but its efficiency is critically dependent on the number of intermediary steps needed for a sufficiently high acceptance rate [3]. Although recent improvements have helped to optimise this technique (most notably the "No-U-turn sampler" introduced in [13]), HMCMC is applicable to only those models for which θ and ξ are continuous quantities. Consequently, it is not well suited to tackle models with discrete variables, *e.g.* disease status [14], or variable dimension number, *e.g.* event data [15].

Another approach is to develop a non-centred parameterisation (NCP) [4, 16, 17] in which a new set of latent variables ξ' (whose distributions are typically independent of the model parameters θ) are introduced. These new variables are related to the original latent variables through some deterministic function $\xi=h(\xi',\theta,y)$. MCMC implemented using this re-parameterisation can lead to improved mixing [17]. In this case, proposed changes to ξ' can be thought of as joint proposals in both ξ and θ.

---

[3] Potentially leading to detrimentally large computational memory requirements.

This paper introduces a new class of MCMC proposal valid for the vast majority of statistical models (Section 2). We refer to these as "posterior-based proposals" (PBPs, Sections 3), as they are constructed with the aid of importance distributions (Section 4) which approximate the posterior. PBPs enable joint updates to both θ and ξ (unlike standard approaches), are fast (*i.e.* they don't require multiple particles like PMCMC), accurate (*i.e.* they draw samples from the true posterior, unlike ABC) and they can be applied to continuous or discrete state space models (unlike HMCMC). A further novelty in PBPs is that they not only account for correlations between θ and ξ inherent to the model, but can also take into account the data (unlike NCP). Application to models used in disciplines ranging from statistical genetics to epidemiology to finance demonstrate that PBPs potentially offer considerable improvements in performance over standard approaches (Section 5).

## 2 Broadly applicable model framework

PBPs are potentially applicable to any statistical model whose conditional dependence structure can be represented by a Direct Acyclic Graph (DAG) [18], as illustrated in Fig. 2. This encompasses a vast range of statistical models including mixed models [19], generalised linear mixed models [20], hidden Markov models [21], discrete time Markov processes [22], and most of the models that can be defined in automated Bayesian software, such as WinBUGS [23], JAGS [24] or Stan [25], which specifically assume a DAG structure. A key property of DAGs is that the indices for the latent variables *e* can be ordered such that each element $\xi_e$ is conditionally dependant on only those other elements with lower index $\xi_{e'<e}$ (a property known as topological ordering)[4].

Simulation results from sequentially sampling each latent variable $\xi_e$ (starting from *e=1* up to *e=E*) from a set of model-defining univariate probability distributions $\pi(\xi_e|\xi_{e'<e},\theta)$. Consequently, the latent process likelihood in Eq.(1) can be expressed as

$$\pi(\xi|\theta) = \prod_{e=1}^{E} \pi(\xi_e | \xi_{e'<e}, \theta). \qquad (2)$$

## 3 Posterior-based proposals (PBPs)

### 3.1 Aim

PBPs first propose a new set of parameters $\theta^p$ relative to $\theta^i$ and then generate $\xi^p$ by means of stochastically *modifying* $\xi^i$ to account for this change in parameters (note, this is in stark contrast to PMCMC which aims to *sample* $\xi^p$ directly from $\pi(\xi|\theta^p,y)$ without reference to $\xi^i$ or $\theta^i$). The novel PBP process involves sequentially sampling each latent variable $\xi_e^p$ from *e=1* up to *E*, and makes use of so-called "importance distributions" (IDs) applied to both the initial and proposed states. These importance distributions $f_{ID}(\xi_e|\xi_{e'<e},\theta,y)$ are approximations to the posterior distributions $\pi(\xi_e|\xi_{e'<e},\theta,y)$ used in importance sampling (see Appendix A for further details). For clarity we leave a discussion of how these approximations are made in practice until Section 4.

### 3.2 Example

We first run through an illustrative example of a PBP and then provide a general description in Section 3.3. Figure 3 shows hypothetical distributions for a particular value of *e* for which $\xi_e$ takes non-negative integer values. Here the black lines represent the true (unknown) distributions for the current $\pi\left(\xi_e \mid \xi^i_{e'<e}, \theta^i, y\right)$ and proposed $\pi\left(\xi_e \mid \xi^p_{e'<e}, \theta^p, y\right)$ states, respectively. Since these curves are approximately Poisson distributed[5], the IDs are taken to be Poisson (as shown by the red dashed lines in Fig. 3) with probability mass functions

$$f_{ID}\left(\xi_e \mid \xi^i_{e'<e}, \theta^i, y\right) = \frac{e^{-\lambda_i} \lambda_i^{\xi_e}}{\xi_e!},$$

$$f_{ID}\left(\xi_e \mid \xi^p_{e'<e}, \theta^p, y\right) = \frac{e^{-\lambda_p} \lambda_p^{\xi_e}}{\xi_e!}. \qquad (3)$$

---

[4] The notation $\xi_{e'<e}$ denotes all elements in ξ with index smaller than *e*.

These functions are characterised by "expected event number" parameters $\lambda_i$ and $\lambda_p$, which themselves are functionally dependent on $(\xi^i_{e'<e}, \theta^i, y)$ and $(\xi^p_{e'<e}, \theta^p, y)$, respectively (see Section 4).

A unique feature of PBPs is that the sampling distribution for $\xi^p_e$ crucially depends on the relative size of $\lambda_p$ and $\lambda_i$. When $\lambda_p > \lambda_i$ (as it is in Fig. 3) $\xi^p_e$ is generated by adding a Poisson distributed variable onto $\xi^i_e$ with expected event number given by the *difference* between $\lambda_p$ and $\lambda_i$:

$$\xi^p_e = \xi^i_e + X \quad \text{where} \quad X \sim Pois\left(\lambda_p - \lambda_i\right) \qquad (4)$$

(*e.g.* in Fig. 3 a random Poisson sample $X=8$ results in $\xi^p_e = \xi^i_e + X = 13$). Such an approach makes sense, because adding an approximately Poisson distributed quantity with expected event number $\lambda_i$ to one with expected event number $\lambda_p - \lambda_i$ gives an approximately Poisson distributed variable with expected event number $\lambda_p$, as required by $\xi^p_e$ on the left hand side of Eq.(4). Thus, Eq.(4) modifies $\xi^i_e$ to generate $\xi^p_e$ accounting for the change in $\lambda$.

On the other hand, when $\lambda_p \leq \lambda_i$ the actual number of events in the proposed state $\xi^p_e$ should be less than in the initial state $\xi^i_e$, because the expected number of events parameter $\lambda$ has reduced. Specifically each existing event in $\xi^i_e$ is retained in $\xi^p_e$ with probability $\lambda_p/\lambda_i$, which is equivalent to sampling $\xi^p_e$ from the following binomial distribution[6]

$$\xi^p_e \sim B(\xi^i_e, \tfrac{\lambda_p}{\lambda_i}). \qquad (5)$$

Together, the two potential sampling schemes for $\xi^p_e$ in Eqs.(4) and (5) (selected depending on whether $\lambda_p$ is bigger or smaller than $\lambda_i$) make up the PBP proposal in cases in which the ID is Poisson. This is summarised by the first line in Table 1.

---

[5] A Poisson distribution expresses the probability a given number of events occur in a fixed interval of time, assuming that events occur at a constant rate.
[6] If $N$ represents the total number of experiments and $p$ is the success probability of each experiment then $B(N,p)$ samples the number of successes.

## 3.3 General approach

In general the choice of ID will depend on the distribution $\pi(\xi|\theta^p,y)$, which itself is model dependent. If, for example, $\pi(\xi_e|\xi_{e'<e},\theta,y)$ is better represented by a normal ID, the PBP sampling procedure is taken from the second line in Table 1. In all, Table 1 summarises sampling schemes for twelve different ID functional forms, each corresponding to probability distributions commonly used in statistical models. These schemes are specifically designed to satisfy the following two conditions:

$$\textbf{Condition 1}: \frac{f_{ID}(\xi_e^p | \xi_{e'<e}^p, \theta^p, y)}{f_{ID}(\xi_e^i | \xi_{e'<e}^i, \theta^i, y)} \frac{g(\xi_e^i)}{g(\xi_e^p)} = 1, \quad (6)$$

$$\textbf{Condition 2}: \xi_e^p \to \xi_e^i \quad \text{as} \quad \theta^p \to \theta^i,$$

where $g(\xi_e^p)$ is the probability of sampling latent variables $\xi_e^p$ starting from state $i$, and $g(\xi_e^i)$ is the probability of sampling $\xi_e^i$ when proposing state $i$ from $p$. Condition 1 ensures that if $\xi_e^i$ is a random sample from $f_{ID}(\xi_e | \xi_{e'<e}^i, \theta^i, y)$ then $\xi_e^p$ will, by construction, be a random sample from $f_{ID}(\xi_e | \xi_{e'<e}^p, \theta^p, y)$ (albeit by design correlated with $\xi_e^i$). Condition 2 guarantees that proposals with small jumps in parameter space have an acceptance probability close to one.

Note, condition 1 can trivially be solved by sampling directly from the IDs

$$\begin{aligned} g(\xi_e^p) &= f_{ID}(\xi_e^p | \xi_{e'<e}^p, \theta^p, y), \\ g(\xi_e^i) &= f_{ID}(\xi_e^i | \xi_{e'<e}^i, \theta^i, y), \end{aligned} \quad (7)$$

but this doesn't satisfy condition 2 and turns out to usually be inefficient[7]. Deriving sampling schemes which also satisfy condition 2 is a non-trivial task guided by intuition and trial and error. Extension of Table 1 to encompass a more comprehensive list of possible sampling distributions will be the subject of future research. The validity of Eqs.(6) for both Poisson and normal IDs is explicitly demonstrated in Appendix B.

---

[7] This approach is akin to standard importance sampling and typically leads to very low acceptance rates for high dimensional models (as observed in examples 5.2 and 5.3 later).

## 3.4 Algorithm

We now describe the general algorithm used to implement PBPs:

---

*POSTERIOR-BASED PROPOSALS*

**Step 1: Generate $\theta^p$** – A proposed set of parameter values is drawn from a multivariate normal (MVN) distribution centred on the current set of parameters in the chain $\theta^i$

$$\theta^p \sim N(\theta^i, j^2 \Sigma^\theta), \qquad (8)$$

where $\Sigma^\theta$ is a numerical approximation to the covariance matrix for $\pi(\theta|y)$ and $j$ is a tuneable jumping parameter (estimation of $\Sigma^\theta$ and optimisation of $j$ are achieved during an initial "adaptation" period, as explained in Appendix C). (Note, performing a joint update on all parameters instead of each individually helps to alleviate poor mixing due to strong parameter correlations in $\pi(\theta|y)$).

**Step 2: Generate $\xi^p$** – We take each latent variable $\xi_e$ in turn (starting from $e=1$ up to $e=E$) and calculate the characteristic quantities defining the IDs for the initial and proposed states (*e.g.* in the Poisson case this would be $\lambda_i$ and $\lambda_p$), as described in Section 4. $\xi_e^p$ is then sampled using specially design proposals outlined in Table 1 (note, this table contains separate lines referring to different potential ID functional forms). This sampling procedure is at the heart of PBPs and represents the key novelty of this approach.

**Step 3: Accept or reject joint proposal for $\theta^p$ and $\xi^p$** – With MH probability (see Appendix D)

$$P_{MH} = \min\left\{1, \frac{\pi(y|\xi^p,\theta^p)\pi(\xi^p|\theta^p)\pi(\theta^p)}{\pi(y|\xi^i,\theta^i)\pi(\xi^i|\theta^i)\pi(\theta^i)} \prod_{e=1}^{E} \frac{f_{ID}(\xi_e^i|\xi_{e'<e}^i,\theta^i,y)}{f_{ID}(\xi_e^p|\xi_{e'<e}^p,\theta^p,y)}\right\}. \qquad (9)$$

---

The algorithm above performs a random walk through the parameter space defined by the posterior, but because the dimensionality of θ is typically much less than ξ, it is expected to mix at a much faster rate[8] than standard MCMC, which performs a random walk in both θ and ξ. Further insights into the PBP procedure are given in Appendix E.

PBPs, by design, result in $\xi^p$ exhibiting correlation with $\xi^i$ (indeed, if $\theta^p=\theta^i$ then $\xi^p$ is exactly the same as $\xi^i$, as required by Eq.(6)). The PBP MCMC algorithm used in this paper mitigates against these correlations by performing PBPs interspersed with standard updates for the latent variables[9] every $U$ steps. Appendix H shows that mixing is not sensitive to the exact value of $U$, and is approximately optimised when $U=4$ (as subsequently used)[10].

---

[8] Subject to sufficiently large jumping size *j* in Eq.(8) being possible.
[9] Which passes through each latent variable and performs a Gibbs or random walk MH update.
[10] In practice the optimum *U* will depend on the relative CPU time needed for PBP and standard updates. If standard updates are much slower it makes sense for *U* to be higher, but typically they are of a similar speed.

# 4 Generating importance distributions (ID)

For each latent variable $\xi_e$, the distribution we wish to approximate is $\pi(\xi_e|\xi_{e'<e},\theta,y)$, *i.e.* the posterior probability distribution for $\xi_e$ given $\xi_{e'<e}$, parameters $\theta$, and data $y$. This distribution can be expressed as

$$\pi(\xi_e | \xi_{e'<e}, \theta, y) = \int\int \pi(\xi_{d\geq e} | \xi_{e'<e}, \theta, y)\, \mathrm{d}\xi_E...\mathrm{d}\xi_{e+1}, \qquad (10)$$

where $\pi(\xi_{d\geq e}|\xi_{e'<e},\theta,y)$ is the joint posterior distribution for latent variables with index $e$ and above (conditional on everything else). The integrals in Eq.(10) successively marginalise over the unknown latent variables, starting with the last $\xi_E$ all the way back to $\xi_{e+1}$. Using Bayes' theorem and Eq.(1), the integrand in Eq.(10) can be expressed as

$$\begin{aligned}\pi(\xi_{d\geq e} | \xi_{e'<e}, \theta, y) &\propto \pi(y|\xi,\theta)\pi(\xi_{d\geq e}|\xi_{e'<e},\theta) \\ &\propto \pi(\xi_e|\xi_{e'<e},\theta)\pi(y|\xi,\theta)\prod_{d=e+1}^{E}\pi(\xi_d|\xi_{e'<d},\theta).\end{aligned} \qquad (11)$$

For now making the simplification that one observation is made per latent variable, *i.e.*

$$\pi(y|\xi,\theta) = \prod_{e=1}^{E}\pi(y_e|\xi_e,\theta), \qquad (12)$$

and substituting Eq.(11) into Eq.(10) gives

$$\begin{aligned}\pi(\xi_e|\xi_{e'<e},\theta,y) &\propto \pi(\xi_e|\xi_{e'<e},\theta)\pi(y_e|\xi_e,\theta) \\ &\times \int\int\prod_{d=e+1}^{E}\pi(\xi_d|\xi_{e'<d},\theta)\pi(y_d|\xi_d,\theta)\,\mathrm{d}\xi_E...\mathrm{d}\xi_{e+1}.\end{aligned} \qquad (13)$$

Analytically performing these integrals is usually not possible. However different levels of approximation can be made depending on the point at which the expression on the right hand side of Eq.(13) is truncated. This leads to a family of importance distributions with increasing accuracy (as illustrated in Fig. 4):

### ID$_0$

By taking just the first term on the right hand side of Eq.(13), the ID is simply set to the distribution from the model itself[11]

$$f_{ID_0}(\xi_e | \xi_{e'<e}, \theta) = \pi(\xi_e|\xi_{e'<e},\theta). \qquad (14)$$

Note, this can only be done provided the model distribution $\pi(\xi_e|\xi_{e'<e},\theta)$ has a functional form belonging to one of the possibilities in Table 1 (or alternatively a new PBP sampling scheme based on the model distribution is created which satisfies the conditions in Eq.(6)). If this cannot be achieved, the functional form for ID$_0$ is chosen to match $\pi(\xi_e|\xi_{e'<e},\theta)$ as closely as possible.

The calculation of ID$_0$, as illustrated in Fig. 4(b), involves only those latent variables on which $\xi_e$ is conditionally dependant. PBPs using ID$_0$ are equivalent to model-based proposals [9] (MBPs), and their MH acceptance probability from Eq.(9) simplifies to

$$P_{MH} = \min\left\{1, \frac{\pi(y|\xi^p,\theta^p)\pi(\theta^p)}{\pi(y|\xi^i,\theta^i)\pi(\theta^i)}\right\}. \tag{15}$$

---

[11] Note, the proportionality sign in Eq.(13) becomes an equality sign because $\pi(\xi_e|\xi_{e'<e},\theta)$ is normalised.

One of the desirable features of MBPs is that they require no hand-tuning. There is a one-to-one correspondence between the model distributions and the proposals, as outlined in Table 1, and so they can be implemented in an automated manner. However in cases in which data substantially restricts model parameters and latent variables, higher order importance distributions become necessary.

## ID$_1$

The ID accuracy is improved by using the first two terms on the right hand side of Eq.(13)

$$f_{ID_1}(\xi_e | \xi_{e'<e},\theta,y) = c\pi(\xi_e | \xi_{e'<e},\theta)\pi(y_e | \xi_e,\theta), \tag{16}$$

where $c$ is a normalising factor. Calculation of ID$_1$, as illustrated in Fig. 4(c), includes not only those latent variables on which $\xi_e$ is dependant, but also the observation $y_e$ on $\xi_e$ itself (as indicated by the green circle)[12].

## ID$_2$

This additionally includes observations on those latent variables which depend on $\xi_e$, e.g. $\xi_{E-1}$ in Fig. 4(d). Equation (13) now gives the improved approximation

$$f_{ID_2}(\xi_e | \xi_{e'<e},\theta,y) = c\pi(\xi_e | \xi_{e'<e},\theta)\pi(y_e | \xi_e,\theta)\int \pi(\xi_{E-1} | \xi_{e'<E-1},\theta)\pi(y_{E-1} | \xi_{E-1},\theta)\mathrm{d}\xi_{E-1}. \tag{17}$$

Higher order approximations (*i.e.* ID$_{n>2}$ which take into account successively more of the model and data[13]) usually prove to be too computationally expensive to be efficient.

## Choosing ID

Choosing which level of ID to optimise PBPs involves a trade-off between the computational cost of generating IDs with the size of posterior jumps (and hence improvement in mixing) they allow. Unfortunately determining *a priori* which option is best is challenging. Indeed, in the results section below we find examples for which ID$_0$, ID$_1$ and ID$_2$ each represent optimum solutions for different problems. From the point of view of the user, the pragmatic approach to take is to first try ID$_0$ (which is the easiest to implement) and if that doesn't help mixing then try ID$_1$ and so on and so forth. Identification of optimal IDs will be the subject of active future research.

The classification scheme presented above is based on models which have one observation per latent variable. For models in which this is not the case, generation of IDs relies on approximating the following expression:

$$\pi(\xi_e | \xi_{e'<e},\theta,y) \propto \pi(\xi_e | \xi_{e'<e},\theta)\int\int \pi(y|\xi,\theta)\prod_{d=e+1}^{E}\pi(\xi_d | \xi_{e'<d},\theta)\,\mathrm{d}\xi_E...\mathrm{d}\xi_{e+1}. \tag{18}$$

---

[12] This relies on the product of model and observation probability distributions being contained within Table 1. If this is not the case then some level of approximation is necessary.
[13] ID$_n$ is the approximation to Eq.(13) in which those latent variables $\xi_d$ that are *n-1* or fewer arrows away from $\xi_e$ (along with any latent variables on which they depend) are included in the integral.

This may or may not be computationally challenging, depending on the scenario considered. However, provided the model itself makes uses of the distributions in Table 1, using $ID_0$ (*i.e.* MBPs) is always possible.

# 5 Empirical evaluation

We now investigate the relative computational performance of PBPs compared to other approaches (where appropriate):

**Standard MCMC** – Here an "update" is performed by sequentially making changes to each model parameter and latent variable in turn. Where possible Gibbs sampling [7] is used, otherwise random walk MH is implemented (note, computational efficiency is optimised by calculating only those parts of the likelihood and observation probability which actually change given a particular proposal).

**Non-centred parameterisation** (NCP) – Standard approaches can also be applied to so-called "non-centred" parameterisations (NCPs). Here inference is performed on (θ,ξ'), where ξ' are distributed independently of θ and the actual latent variables are related through deterministic relationship ξ=h(ξ',θ,y) [26]. To give a simple example, suppose each latent variable is distributed normally $ξ_e \sim N(μ, σ^2)$ with mean $μ$ and variance $σ^2$ being model parameters θ. This can be reparametrized by setting $ξ'_e \sim N(0,1)$, with the functional dependency $h$ being given by $ξ_e = μ + σξ'_e$. Note, NCPs are not always possible because parameters cannot always be separated from distributions, (*e.g.* this cannot be done for the Bernoulli, Poisson or Gamma distributions), hence NCPs cannot be used in examples 5.1 and 5.3 below). More complicated schemes which make use of partial CP/NCP proposals and interweaving different parameterisations [4, 16, 17, 27] are not considered here.

**Hamiltonian MCMC (HMCMC)** – This generates samples by integrating a trajectory from the current parameter and latent variable state to a proposed state (typically via many intermediary step) [3, 12]. Such a process accounts for gradients in the log of the posterior probability, allowing large distances in parameter and latent variable space to be traversed. Optimisation balances making the initial and proposed states as uncorrelated as possible (to improve mixing), whilst reducing the computational burden of excessive steps and allowing for a sufficiently good acceptance probability. The Metropolis-adjusted Langevin algorithm [28] (MALA) is a special case of HMCMC in which only a single step is taken. HMCMC is limited to only those models with continuous model parameters and latent variables, and consequently cannot be applied to examples 5.1 and 5.4 below.

**Particle MCMC (PMCMC)** – This approach generates unbiased approximations $\hat{π}(y|θ)$ by means of a sequential filtering process [2]. In its simplest implementation this involves running multiple simulations of the model (*i.e.* sampling from $ID_0$ from Section 4) which are periodically filtered based on the observations, however more efficient schemes can make use of higher order importance distributions $ID_1$, $ID_2$, *etc...* Jumping in parameter space (*e.g.* using Eq.(8)) is achieved through a MH algorithm that makes uses of these unbiased estimates.

Further details, along with optimisation procedures for each of the different methods, are described in Appendices I-K.

Additionally, it should be pointed out that another technique to improve mixing is to simply integrate out problem parameters directly. Such an approach, however, is not considered in this paper for two reasons: firstly it lacks generality, because it only applies to models in which these integrals can actually be performed, and secondly it restricts the possible priors that can be applied to a given model (most prior choices make such integration impossible).

Four contrasting model types are investigated. In each case we examine the efficiency of the various algorithms for parameter inference from data simulated from the models in question. Results shown are based on $10^6$ MCMC updates preceded by $10^4$ discarded samples from a burn-in/adaptation period (see Appendix C). One way to measure MCMC efficiency is to calculate the "effective sample size" [1] (see Appendix L), and here we calculate the computational time for any given algorithm to generate 100 effective posterior samples[14] of a given parameter[15]. Unless otherwise stated, uninformative flat priors are assumed.

## 5.1 Inferring disease prevalence and diagnostic test performance

Suppose we aim to estimate the disease prevalence (fraction of infected) $p_D$ in a population of $P$ individuals using cross-sectional diagnostic test results. Such diagnostic tests are typically imperfect, and characterised by a sensitivity *Se* (the probability of a positive test result given an infected individual) and specificity *Sp* (the probability of a negative test result given uninfected). Suppose *Se* and *Sp* are unknown. In the absence of a gold standard defining which individuals are truly infected, inference is only possible when two or more independent test results are recorded per individual (due to confounding). Here we assume that results are available from a single diagnostic test performed on each individual at two times, labelled $t=\{1,2\}$. The model is shown in Fig. 5(a) and described in detail along with development of PBP proposals in Appendix M. We note that this model could be embedded in more complex models, *e.g.* fitted to data from capture-mark-recapture programmes [29].

**Speed comparison**

Simulated data was created using $p_D$=0.5, $Se_1$=$Se_2$=0.6 and $Sp_1$=$Sp_2$=0.9 for $P$=1000 individuals. Inference was then performed using MBP/PBP MCMC approaches as well as standard Gibbs sampling. Figure 5(b) shows how posterior samples for $p_D$ vary as each of the three algorithms progress. By binning these samples, the marginal posterior distributions for $p_D$ can be generated, as shown in Fig. 5(c) (which also shows results for the other model parameters). These distributions contain the known values used to generate the data (denoted by vertical black lines), indicating successful inference.

It is important to note that in the limit of infinite MCMC sample number all algorithms generate exactly the same set of marginal distributions (*i.e.* those shown in Fig. 5(c)). However the speed with which they converge *does* vary. For example, Fig. 5(d) shows how the computational time (*i.e.* the CPU time to generate 100 effective samples) to infer $p_D$ increases with population size. Here we find that MBPs (ID$_0$) actually performs worse than the standard Gibbs approach, however when observations are incorporated (ID$_1$) the resultant PBP algorithm is around three times faster than Gibbs sampling (at least when the number of individuals is large). Although mixing is greatly improved, as is evident in comparing Fig. 5(b) and Fig. 5(c), this is offset by the computational overhead associated with each PBP.

---

[14] Note, all MCMC chains were run long enough to be well mixed, with ESS typically exceeding 1000 and at least greater than 500.

[15] Simulation and inference were averaged over twenty separate runs to help remove data-dependent noise.

PMCMC which use $ID_0$ (*i.e.* the usual approach of simulating from the model) are found to perform very poorly (this is because a very large number of particles are required for a reasonable acceptance rate). However when $ID_1$ is used, PMCMC actually becomes the fastest approach. In this particularly simple example, $ID_1$ exactly represents the posterior and so PMCMC only requires a single particle to run. This is atypical, and usually particle methods fail when large numbers of observations are made (as demonstrated later). Because this model contains discrete latent variables (*i.e.* the underlying disease status of individuals $D_e$ are 0 or 1), HMCMC approaches are not possible, and the Bernoulli distribution does not allow for NCP.

This example demonstrates a possible modest improvement in computational speed when using PBPs compared to a standard Gibbs approach. The next example, however, shows that much larger potential gains can be made.

## 5.2 Stochastic volatility model

Stochastic volatility (SV) models are used to capture time-varying volatility on financial markets, and are essential tools in risk management, asset pricing and asset allocation [30]. In economics, a "logarithmic rate of return" can be defined by $y_e$=log($V_{e+1}/V_e$), where $V_e$ is the price of an asset (*e.g.* a share) on day *e*. Consequently, when $y_e$ is positive it means that on day *e+1* the asset goes up in price, but when it is negative it goes down. One way to capture time variation in $y_e$ is through the so-called SV*t* model [31]

$$\begin{aligned} y_e &= e^{h_e/2} u_e, \\ h_e &= \mu + \phi(h_{e-1} - \mu) + \eta_e, \end{aligned} \quad (19)$$

where $u_e$ are i.i.d. with a Student's *t*-distribution (characterised by parameter ν) and $\eta_e$ are i.i.d. normal (with zero mean and variance $\sigma^2$). Note, if $h_e$ is fixed (*i.e.* σ=0), the logarithmic rate of return $y_e$ would simply be sampled asymptotically from a distribution with fixed variance, or "volatility". The introduction of time variation in the variable $h_e$, whose temporal correlations are measured by 0<$\phi$<1, means that $y_e$ experiences stochastic volatility, *i.e.* periods when there are large variations in asset price, and other periods when there isn't much variation.

The SV*t* model is represented in Fig. 6(a). Simulated data was created using $\mu$=-10, $\phi$=0.99, ν=12, $\sigma^2$=0.0121 (which are parameter values based on estimates made from the S&P 500 index [30]). Figure 6(b) shows the time variation in $y_e$ and $h_e$ (observe how changes in $h_e$ correspond to changes in the volatility of $y_e$) over *E*=3000 days. A detailed development of PBP proposals for this model is given in Appendix N.

### Speed comparison
Bayesian inference from the simulated data in Fig. 6(c) identifies marginal posterior distributions which contain the true parameter values, as indicated by black vertical lines. For each algorithm Fig. 6(d) shows how the computational time to infer $\sigma^2$ varies with the correlation parameter $\phi$ used to generate the data (with all other parameters fixed as above).

The standard algorithm is at its fastest when $\phi\sim1$ but slows down considerably as $\phi$ is reduced. Some speed up is observed when NCP is used (where $h'_e = (h_e - \mu)/\sigma$), however the MBP and PBP methods are found to be much faster (note PBP using $ID_2$ was not found to be any faster than with $ID_1$, and so is not shown). HMCMC was found to be relatively slow using the standard parameterisation, but markedly increased in speed when using NCP. PMCMC methods (using either $ID_0$ or $ID_1$) were found to be extremely slow (because they required a huge number of particles), and lie above the top of this graph.

For real financial markets $\phi$ is within the range 0.95 to 0.99 [30], reflecting a high degree of persistence in volatility. This corresponds to PBPs running between two and four times faster than the standard approach, and comparable in speed with HMCMC (sometimes faster and sometimes slower)[16]. However, the left hand side of Fig. 6(d) clearly demonstrates the existence of regimes for which PBPs are faster by a factor exceeding 10 than all other methods tried.

## 5.3 Mixed model

Mixed models (MMs) [32] explain observations in terms of both "fixed effects" (*e.g.* individual attributes such as gender or disease status) and "random effects" (which account for random uncontrollable factors within a study, *e.g.* variation in student grades as a result of variation in the quality of schools). MMs are useful in a wide variety of applications in the physical, biological and social sciences [33-36]. They assume that a vector of *N* measurements *y* can be decomposed into three contributions:

$$y = X\beta + Za + \varepsilon, \qquad (20)$$

where **X** and **Z** are design matrices that define model structure, $\beta$ is a vector of *F* fixed effects, **a** is a vector of *E* random effects and $\varepsilon$ are residuals. Random effects and residuals are assumed to be MVN with zero mean and covariance matrices **G** and **R**, respectively. For simplicity we assume that $G = \sigma_a^2 A$, where **A** is a known symmetric matrix, and $R = \sigma_\varepsilon^2 I$, where **I** is the identity matrix (such that residuals are uncorrelated between measurements). The quantity $r^2 = \sigma_a^2/(\sigma_a^2 + \sigma_\varepsilon^2)$ measures the relative contribution of random effects to residuals.

**Speed comparison**

An important application of mixed models is in the field of quantitative genetics, which aims to understand the genetic basis of traits of interest [37]. As an illustration take *y* to represent measurements of height within a population. These measurement are correlated, *e.g.* if an individual has tall parents they are more likely to be tall. This correlation results from genetic inheritance. The relatedness of individuals within the population is captured by the so-called "relationship matrix" **A**.

Simulated data were generated assuming a population randomly mated over four generations with $N=E=4\times10^3$ individuals and $F=2$ fixed effects (see Appendix O for further details). Figure 7(a) shows how the computational time to infer $r^2$ varies with the $r^2$ value used to generate the data. Disregarding fixed effects, $r^2=0$ implies no genetic inheritance and $r^2=1$ corresponds to a trait dominated by genetic inheritance. In these limits standard Gibbs sampling slows down significantly due to strong parameter-latent variable correlations in the posterior, leading to poor mixing. Using MBPs with $ID_0$ shows an improvement over Gibbs for low $r^2$, however for high $r^2$ its performance proves to be poor. Using $ID_1$ and $ID_2$ leads to further improvements in computational speed, resulting in PBPs becoming consistently faster than the standard Gibbs approach (*e.g.*, $ID_2$ is a factor ~50 times faster in the limit $r^2 \rightarrow 0$). The move to $ID_3$ shows little improvement and for $ID_4$, $ID_5$ *etc.* PBPs become progressively slower despite mixing better. Consequently $ID_2$, which uses measurements taken on individuals as well as close relatives, represents an optimum choice in this particular example.

---

[16] The reason PBPs become slower as $\phi \to 1$ is that (due to correlations introduced by $\phi$) $\pi(\xi_e|\xi_{e'<e},\theta,y)$ is expected to depend on observations up to around $1/(1-\phi)$ days ahead. In contrast, $ID_n$ only includes observations up to *n*-1 days ahead, which is typically a much shorter interval.

Results for HMCMC in Fig. 7(b) are very slow. The reason lies in the fact that the trajectories themselves are found to behave diffusively (*e.g.* if a trace plot of $\sigma_a^2$ is made, its path exhibits familiar random walk behaviour, rather than the relatively smooth progress from one side of the posterior to the other that would be hoped for). NCP HMCMC (which sets ***a'= a**/σ_a*) led to a marked improvement, but still it remained considerably slower than the other methods.

One of the striking features of Fig. 7(b) is how well the NCP standard approach works (only around two times slower than the best PBPs for lower $r^2$). The reason is that for this particular model NCP and MBPs work in much the same way: under NCP, proposals in $\sigma_a$ lead to a simultaneous expansion or contraction of the random effects (through reparameterization ***a**= σ_a **a'***), and this is also what happens in MBPs when the value of *κ* in Table 1 is set to zero (note, the two curves for these methods lie very close to each other in Fig. 7(b), with NCP slightly faster due to the fact that fewer computations are required per update).

PBP and NCP approaches have also been applied to mixed models and generalised mixed models (binary disease data) with diagonal, sparse and dense ***A*** (results not shown) and overall PBPs were not found to perform substantially faster than NCP, and in some cases were slower. Consequently for these types of model, standard approaches using NCP may prove to be the best method to use, particularly given the relative ease with which they can be implemented (Table 2).

## 5.4 Logistic population model

We here consider a simple illustrative example taken from ecology. Imagine time is discretised in intervals of size τ and suppose we are following the population size of an animal species which has been released into the wild. We know that births and deaths will occur, and that the population size will increase, but that increase will be curtailed by the limitation of resources within the area. This can be modelled in the following way

$$\begin{aligned} b_t &\sim \text{Pois}(\tau r_b P_t(1-P_t/K)), \\ d_t &\sim \text{Pois}(\tau \mu P_t), \\ P_{t+1} &= P_t + b_t - d_t, \end{aligned} \quad (21)$$

where $b_t$, $d_t$ and $P_t$ are the number of births, deaths and population size in time interval *t*, Pois(λ) generates Poisson distributed integer samples with mean λ, $r_b$ is the birth rate, $\mu$ is the mortality and *K* is the carry capacity (which determines the maximum size of the population). Equation (21) can be considered as a Tau-leaping approximation to the underlying continuous time process under study [38].

A DAG for this model is shown in Fig. 8(a) and a simulation is shown in Fig. 8(b). Now suppose that to keep track of this wildlife population traps are set at certain points in time and the number of trapped individuals is recorded (shown by the red crosses)[17]. Note, animals are caught with capture probability *p*, and so these result are much less that the actual population sizes and also contain additional stochastic noise.

The data from Fig. 8(b) alone is insufficient to estimate all four model parameters, so here semi-informative priors are placed on the mortality rate $\mu$ and capture probability *p* (in reality captured animals are marked and then when re-capture this provides direct evidence for these quantities)[18]. Figure 8(c) shows the results of inference, with $\mu$ and *p* largely following their prior distributions and reasonably good estimates being obtained for birth rate $r_b$ and carrying capacity *K*.

### Speed comparison

The CPU time to estimate 100 effective samples of $r_b$ is shown in Fig. 8(d) as a function of the number of measurements made during the time interval. On the right hand side is the extreme case in which measurements are made at every single time point. Here the observations tightly restrict the potential values for the latent variables, and the standard approach is actually found to perform the best. On the other hand as fewer and fewer observations are made, both the MBP and PMCMC methods become more and more efficient. Note, however that MBPs are consistently around 5 times faster than PMCMC. Despite mixing faster, PMCMC take much longer per update because of the large number of particles (simulations of the process) needed.

Unfortunately here PBPs are not found to be effective because identification of the importance distribution at time *t* is informed by the next measurement, which is potentially many time steps into the future. Further work is need to develop effective IDs in these types of situation, which go beyond the simple classification scheme from Section 4.

Note, HMCMC were not possible because of the discrete latent variables in the model and NCP methods could not be used because model parameters cannot be separated from Poisson distributed latent variables.

## 6 Summary

This paper introduced posterior-based proposals (PBP) and demonstrated that they speed up MCMC inference in many cases where existing approaches perform poorly. PBP are applicable to the majority of statistical models (namely, those whose conditional dependence structure is expressible in terms of a directed acyclic graph). Performance is enhanced by improving mixing of MCMC (*i.e.* increasing the rate at which they generate uncorrelated samples) by jointly proposing changes to model parameters θ and latent variables ξ (see Section 3 for details). PBPs are a family of proposal schemes built by generating importance distributions ($ID_0$, $ID_1$, *etc*.) that systematically account for dependence structure in the Bayesian posterior with increasing accuracy. The zeroth-order approximation $ID_0$ ignores the data and thus corresponds to "model-based proposals" (MBPs) [9]. The optimal level of approximation depends on problem specific trade-offs between improved mixing resulting from increased acceptance rates and the computational cost of generating suitably accurate IDs.

---

[17] Trapped animals are then released back into the wild.
[18] Specifically a gamma distributed prior on $\mu$ with mean 0.3 and variance 0.0144 and a beta distributed prior on *p* with mean 0.5 and variance 0.0025.

The relative computational speed of PBPs compared to "standard" Gibbs/random walk MH techniques (using centred and non-centred parameterizations) as well as HMCMC and PMCMC approaches was investigated for various benchmark models used in applications ranging from finance to ecology to statistical genetics. Whilst different methods worked better or worse in different scenarios, PBPs were found to be either near to the fastest or significantly faster than the next best approach (by up to a factor of 10). Table 2 summarizes the relative strengths and weaknesses of the various approaches.

# 7 Discussion

We now discuss PBPs in relation to PMCMC and HMCMC under two regimes: "model dominant" and "data dominant". Here we define "model dominant" to relate to problems in which the shape of the posterior is largely represented by the latent process likelihood, with the observation model providing a small perturbation on top of this. A good example of this would be a complex model applied to relatively few actual measurements (*e.g.* the left hand side of Fig. 8(d)). The flip side of this argument, however, is the "data dominant" regime, in which the data exceeds or is comparable to the model complexity[19] (*e.g.* the right hand side of Fig. 8(d)).

As stated previously, PMCMC only works on problems in which data can be incorporated in a sequential manner. Even then, however, PMCMC can become slow in the data dominant regime due to requiring a very large number of particles to give a reasonable acceptance rate[20]. This was explicitly demonstrated in examples 5.2 and 5.3 above, where PMCMC was found to be vastly slower than the other approaches. However, a key advantage of PMCMC approaches is that they are usually relatively easy to implement and typically allow for efficient parallelization.

HMCMC relies on calculating gradients in the log-likelihood, and hence is not applicable to models with discrete variables (*e.g.* Sections 5.1 and 5.4) or when the number of variables within the model changes. Both of these challenges are frequently encountered in models in which a latent variable represents some unknown state of the system such as disease state or other individual classification. HMCMC efficiency is not related to whether a given problem is model or data dominant, but is very much dependent on the specific shape of the posterior itself. Often it is tested on high dimensional multivariate normal-type posterior distributions, where it is found to perform well against other approaches [3]. However in many real-world problems, Hamiltonian trajectories can suffer from random walk-type behaviour (as was observed in example 5.3) as a result of parameter/latent variable correlations. These trajectories necessitate a large number of small intermediary steps for each MCMC update, significantly reducing algorithm performance. In contrast to PMCMC, HMCMC cannot easily be parallelised.

Finally we come to PBPs. Generally speaking they tend to work better (in comparison to both standard approaches and HMCMC) on problems which are model dominant. The reason can be seen if we look at the limiting case in which there is no data. Here PBPs (which are actually MBPs in this particular case) easily map out the prior distribution for model parameters (indeed Eq.(15) shows that the PBP algorithm generates random walk MH behaviour in parameter space with acceptance probability given simply by $P_{MH}=\min\{1,\pi(\theta^p)/\pi(\theta^i)\}$). The introduction of importance distributions in Section 4 allow for the data itself to be incorporated into the proposals, so helping maintain the efficiency of PBPs as they move out of the model dominated regime towards the data dominated case.

---

[19] Which could be measured by comparing the number of data points and the number of model parameter and latent variables.
[20] This happens in cases in which after simulating between successive time points the probability of the observed data is small, either because the observations themselves are very specific, or because multiple measurements are made.

The main challenges facing PBPs are twofold: firstly the development of fast and accurate importance distributions that help PBPs in the data dominant regime (*e.g.* on the right hand side of Fig. 8(d)) and secondly the identification of PBP proposals for a broad range of distributions (*i.e.* extending Table 1 to include other distributions). Like HMCMC, PBPs also cannot easily be parallelised. However a potential future extension to PBPs would be to incorporate them into a particle-like framework in which multiple proposals are made along with a particle filter applied at different time points. This may lead to further improvements in speed under some scenarios, and will be the subject of future investigation.

One final point to mention is that whilst some complex models may not fit into the DAG structure required by PBPs, it may be that certain subsections of them do. Here PBPs could still profitably be used by fixing all non-DAG elements of the model under the proposal (whereby they are incorporated into the data $y$ for the purposes of the proposal step).

The introduction of PBPs offers a promising opportunity for optimising MCMC and for further improvements, *e.g.* through creating particle versions of PBPs and the development and automation of efficient importance distributions under different scenarios. As such we believe PBPs are an exciting new methodology which will complement other tools in the MCMC toolbox.

## Data accessibility
All of the C++ computer codes used to generate the results in Section 5 are included in the supplementary material.

## Competing interests
We have no competing interests.

## Authors' contributions
CMP introduced the ideas behind PBPs and obtained the results, SCB introduced application of PBPs to quantitative genetics models, ADW help clarify description of the methodology, and GM helped develop the manuscript along with contributing to the methodology itself. All authors gave final approval for publication.

## Funding statement
This research was funded by the Strategic Research programme of the Scottish Government's Rural and Environment Science and Analytical Services Division (RESAS). ADW's contribution was funded by the BBSRC Institute Strategic Programme Grants.

# Tables

| ID | Prob. Dist. | Posterior-based proposal | |
|---|---|---|---|
| Poisson $\xi_e \sim Pois(\lambda)$ | $\dfrac{\lambda^{\xi_e} e^{-\lambda}}{\xi_e !}$ | $\lambda_p > \lambda_i$ | $\xi_e^p = \xi_e^i + X$ where $X \sim Pois(\lambda_p - \lambda_i)$ |
| | | $\lambda_p \leq \lambda_i$ | $\xi_e^p \sim B(\xi_e^i, \tfrac{\lambda_p}{\lambda_i})$ |
| Normal $\xi_e \sim N(\mu, \sigma^2)$ | $\dfrac{1}{\sqrt{2\pi\sigma^2}} e^{-\tfrac{(\xi_e-\mu)^2}{2\sigma^2}}$ | $\sigma_p > \sigma_i$ | $\xi_e^p \sim N\!\left(\mu_p + \alpha(\xi_e^i - \mu_i), \kappa(\sigma_p^2 - \sigma_i^2)\right)$ where $\alpha^2 = \kappa + (1-\kappa)\tfrac{\sigma_p^2}{\sigma_i^2}$ |
| | | $\sigma_p \leq \sigma_i$ | $\xi_e^p \sim N\!\left(\mu_p + \alpha \tfrac{\sigma_p^2}{\sigma_i^2}(\xi_e^i - \mu_i), \kappa \tfrac{\sigma_p^2}{\sigma_i^2}(\sigma_i^2 - \sigma_p^2)\right)$ where $\alpha^2 = \kappa + (1-\kappa)\tfrac{\sigma_i^2}{\sigma_p^2}$. $\kappa$ is a tuneable constant (typically set to 0.03, as discussed in Appendix F). |
| Exponential $\xi_e \sim Exp(r)$ | $r e^{-r\xi_e}$ | $r_p > r_i$ | $X \sim Exp(r_p - r_i)$, if $X > \xi_e^i$ then $\xi_e^p = X$ else $\xi_e^p = \xi_e^i$ |
| | | $r_p \leq r_i$ | $u$ is a random number between 0 and 1. if $u < 1 - \tfrac{r_p}{r_i}$ then $\xi_e^p = \xi_e^i + X$, $X \sim Exp(r_p)$ else $\xi_e^p = \xi_e^i$ |
| Gamma $\xi_e \sim \Gamma(\alpha, \beta)$ | $\dfrac{\beta^\alpha \xi_e^{\alpha-1} e^{-\beta\xi_e}}{\Gamma(\alpha)}$ | $\alpha_p > \alpha_i$ | $\xi_e^p = \tfrac{\beta_i}{\beta_p}\left(\xi_e^i + X\right)$ where $X \sim \Gamma(\alpha_p - \alpha_i, \beta_i)$ |
| | | $\alpha_p \leq \alpha_i$ | $\xi_e^p = \tfrac{\beta_i}{\beta_p} \xi_e^i Y$ where $Y \sim Beta(\alpha_p, \alpha_i - \alpha_p)$ |
| Beta $\xi_e \sim Beta(\alpha, \beta)$ | $\dfrac{\Gamma(\alpha+\beta)}{\Gamma(\alpha)\Gamma(\beta)} \xi_e^{\alpha-1}(1-\xi_e)^{\beta-1}$ | $\alpha_p > \alpha_i$, $\beta_p = \beta_i$ | $\xi_e^p = 1 - (1 - \xi_e^i) X$ where $X \sim Beta(\beta_i + \alpha_i, \alpha_p - \alpha_i)$ |
| | | $\alpha_p \leq \alpha_i$, $\beta_p = \beta_i$ | $\xi_e^p = \dfrac{\xi_e^i X}{1 - \xi_e^i(1 - X)}$ where $X \sim Beta(\alpha_p, \alpha_i - \alpha_p)$ |
| | | $\beta_p > \beta_i$, $\alpha_p = \alpha_i$ | $\xi_e^p = \xi_e^i X$ where $X \sim Beta(\alpha_i + \beta_i, \beta_p - \beta_i)$ |
| | | $\beta_p \leq \beta_i$, $\alpha_p = \alpha_i$ | $\xi_e^p = \dfrac{\xi_e^i}{\xi_e^i + (1 - \xi_e^i) X}$ where $X \sim Beta(\beta_p, \beta_i - \beta_p)$ |
| Bernoulli $\xi_e \sim Bern(z)$ | $\Pr(\xi_e = 1) = z$, $\Pr(\xi_e = 0) = 1 - z$ | $z_p > z_i$ | if $\xi_e^i = 1$ set $\xi_e^p = 1$ else $\left\{ \text{if } u < \tfrac{z_p - z_i}{1 - z_i} \text{ then set } \xi_e^p = 1 \text{ else } \xi_e^p = 0 \right\}$ where $u$ is a random number between 0 and 1 |
| | | $z_p \leq z_i$ | if $\xi_e^i = 0$ set $\xi_e^p = 0$ else $\left\{ \text{if } u < 1 - \tfrac{z_p}{z_i} \text{ then set } \xi_e^p = 0 \text{ else } \xi_e^p = 1 \right\}$ where $u$ is a random number between 0 and 1 |

**Table 1(a):** Shows how to sample the proposed latent variable $\xi_e^p$ from $\xi_e^i$ given some importance distribution (ID) functional form (left hand column), whose characteristic parameters change from some initial set to a proposed set.

| ID | Prob. Dist. | Posterior-based proposal | |
|---|---|---|---|
| Binomial $\xi_e \sim B(N,z)$ | $C_{\xi_e}^N z^{\xi_e}(1-z)^{N-\xi_e}$ where $C_n^N = \frac{N!}{n!(N-n)!}$ | $z_p > z_i$, $N_p = N_i$ | $\xi_e^p = \xi_e^i + X$ where $X \sim B(N_i - \xi_e^i, \frac{z_p - z_i}{1-z_i})$ |
| | | $z_p \leq z_i$, $N_p = N_i$ | $\xi_e^p \sim B(\xi_e^i, \frac{z_p}{z_i})$ |
| | | $N_p > N_i$, $z_p = z_i$ | $\xi_e^p = \xi_e^i + X$ where $X \sim B(N_p - N_i, z_i)$ |
| | | $N_p \leq N_i$, $z_p = z_i$ | $\xi_e^p \sim HG(N_i, N_p, \xi_e^i)$ |
| Uniform $\xi_e \sim Uni(a,b)$ | $\left.\begin{array}{l}\frac{1}{b-a}\\ 0\end{array}\right\} \begin{array}{l} a \leq \xi_e \leq b \\ \xi_e < a \text{ or } \xi_e > b\end{array}$ | | $\xi_e^p = a_p + \left(\frac{\xi_e^i - a_i}{b_i - a_i}\right)(b_p - a_p)$ |
| Geometric $\xi_e \sim Geom(z)$ | $(1-z)^{\xi_e} z$ | $z_p > z_i$ | $X \sim Geom(\frac{z_p - z_i}{1-z_i})$, if $X < \xi_e^i$ then $\xi_e^p = X$ else $\xi_e^p = \xi_e^i$ |
| | | $z_p \leq z_i$ | $u$ is a random number between 0 and 1 if $u < 1 - \frac{z_p}{z_i}$ then $\xi_e^p = \xi_e^i + 1 + X$, $X \sim Geom(z_p)$ else $\xi_e^p = \xi_e^i$ |
| Negative Binomial $\xi_e \sim NB(r,z)$ | $C_{\xi_e}^{\xi_e + r - 1} z^{\xi_e}(1-z)^r$ where $C_n^N = \frac{N!}{n!(N-n)!}$ | $z_p > z_i$, $r_p = r_i$ | $\xi_e^p = \xi_e^i + q + X$ where $q \sim B(r_i, \frac{z_p - z_i}{1-z_i})$, $X \sim NB(q, z_p)$ |
| | | $z_p \leq z_i$, $r_p = r_i$ | $\xi_e^p \sim X$ where $X$ is drawn from p.m.f. $\sum_{q=1}^{r} \left( C_q^r C_{\xi_e^i - X - q}^{\xi_e^i - X - 1} C_X^{X+r-1} / C_{\xi_e^i}^{\xi_e^i + r - 1} \right)\left(1 - \frac{z_p}{z_i}\right)^q \left(\frac{z_p}{z_i}\right)^X$ |
| | | $r_p > r_i$, $z_p = z_i$ | $\xi_e^p = \xi_e^i + X$ where $X \sim NB(r_p - r_i, z_i)$ |
| | | $r_p \leq r_i$, $z_p = z_i$ | $\xi_e^p \sim NHG(\xi_e^i + r_i - 1, \xi_e^i, r_p)$ |
| Lognormal $\xi_e \sim \text{lnorm}(\mu, \sigma^2)$ | $\frac{1}{\xi_e \sqrt{2\pi\sigma^2}} e^{-\frac{(\log \xi_e - \mu)^2}{2\sigma^2}}$ | $\sigma_p > \sigma_i$ | $\log \xi_e^p \sim N\left(\mu_p + \alpha(\log \xi_e^i - \mu_i), \kappa(\sigma_p^2 - \sigma_i^2)\right)$ where $\alpha^2 = \kappa + (1-\kappa)\frac{\sigma_p^2}{\sigma_i^2}$ |
| | | $\sigma_p \leq \sigma_i$ | $\log \xi_e^p \sim N\left(\mu_p + \alpha \frac{\sigma_p^2}{\sigma_i^2}(\log \xi_e^i - \mu_i), \kappa \frac{\sigma_p^2}{\sigma_i^2}(\sigma_i^2 - \sigma_p^2)\right)$ where $\alpha^2 = \kappa + (1-\kappa)\frac{\sigma_i^2}{\sigma_p^2}$ |
| Logistic $\xi_e \sim \text{logistic}(\mu, s)$ | $\frac{e^{-\frac{\xi_e - \mu}{s}}}{s\left(1 + e^{-\frac{\xi_e - \mu}{s}}\right)}$ | | $\xi_e^p = \mu_p + (\xi_e^i - \mu_p)\frac{s_p}{s_i}$ |

**Table 1(b):** Here draws from the hypergeometric distribution $X \sim HG(N, K, n)$ have p.m.f. $C_X^K C_{n-X}^{N-K} / C_n^N$ and from the negative hypergeometric distribution $X \sim NHG(N, K, r)$ have p.m.f. $C_X^{X+r-1} C_{K-X}^{N-r-X} / C_K^N$.

| Method | Description | Performance | Limitation | Optimisation/ Implementation |
|---|---|---|---|---|
| MBP | Generate $\theta^p$ from $\theta^i$ and modify $\xi^i$ to generate $\xi^p$ (based on the model). | Found to exhibit faster mixing than standard approach in a large number of scenarios, particularly those which are "model dominant". | Requires that the distributions used in the model are found in Table 1 (or can be derived using the conditions in Eq.(6)). | **Easy** – The only free parameter is $U$ (which governs the relative rate of MBP to standard updates), and results are found to be relatively insensitive to its value. |
| PBP | Generate $\theta^p$ from $\theta^i$ and modify $\xi^i$ to generate $\xi^p$ (based on importance distributions that account for the data). | Found to be the fastest approach in many cases. Provides additional computational performance compared to MBPs, especially in the "data dominant" regime. | Requires that the importance distributions used are found in Table 1 (or can be derived using the conditions in Eq.(6)). | **Medium-Hard** – The most challenging aspect of PBPs is obtaining good importance distributions that approximate the posterior. In the case of ID$_1$ this is often relatively straightforward, but higher order approximations can be more difficult, if not impossible, to identify. |
| HMCMC | Randomly sample momentum vector at $(\theta^i,\xi^i)$ and integrate Hamilton's equations to generate $(\theta^p,\xi^p)$. | Can work very well under certain circumstances (*e.g.* it was the fastest approach for a small parameter region in Fig. 6(d)), however can be slow if integrated trajectories behave diffusively (which results in a very large number of steps being necessary to pass from one side of the posterior to the other). | Only works with continuous parameters and latent variables and in situations in which the gradients in the log-likelihood can be calculated. | **Hard** – Optimising the step size is relatively easy, as it can be selected to achieve a certain acceptance rate. Optimising the number of steps for each update, however, is difficult and efficiency is found to critically depend on this value. Automated methods such have NUTs have been developed, but these are not easy to implement. Also choice of CP/NCP was found to substantially affect performance. |
| PMCMC | Generate $\theta^p$ from $\theta^i$ and use particles to generate unbiased estimate $\hat{\pi}(y\|\theta)$. This is then used in a MH update. | These were found to work reasonably well in the "model dominant" regime. In the idealised case when latent variable can be exactly sampled (requiring only a single particle) they are fastest, but in the other examples investigated MBPs/PBPs were found to outperform. | Observations on the system need to be sequentially ordered (such as in time series data) to allow for particle filtering to work. | **Easy-Hard** – Simulation from the model and subsequent particle filtering usually straightforward. However, things are more complicated when importance distribution are required (to avoid an unreasonably large number of particles), similar to the challenges faced by PBPs. |
| Stand. / NCP | Make local changes to θ and ξ. Gibbs sampling is performed where possible, otherwise random-walk MH updates. | Standard approaches tend to work well in the "data dominant" regime (here parameter and latent variable values are well established and correlations between them are less important. Using NCP, where possible, was often found to substantially increase speed. | None | **Easy-Medium** – When performing local updates, care is needed to only calculate those parts of the likelihood and observation model which change. Often CP and NCP work in different regimes, hence overall optimisation would require performing combinations of the two. |

**Table 2:** This table gives a brief description of the methods along with various pros and cons. Here "data dominant" refers to situations in which the number of observations is similar (or exceeding) the number of latent variables and "model dominant" relates to the converse case in which few observations are made on a model containing many latent variables. MH stands for Metropolis-Hastings, and CP and NCP stand for centred and non-centred parameterisations.

# Figures

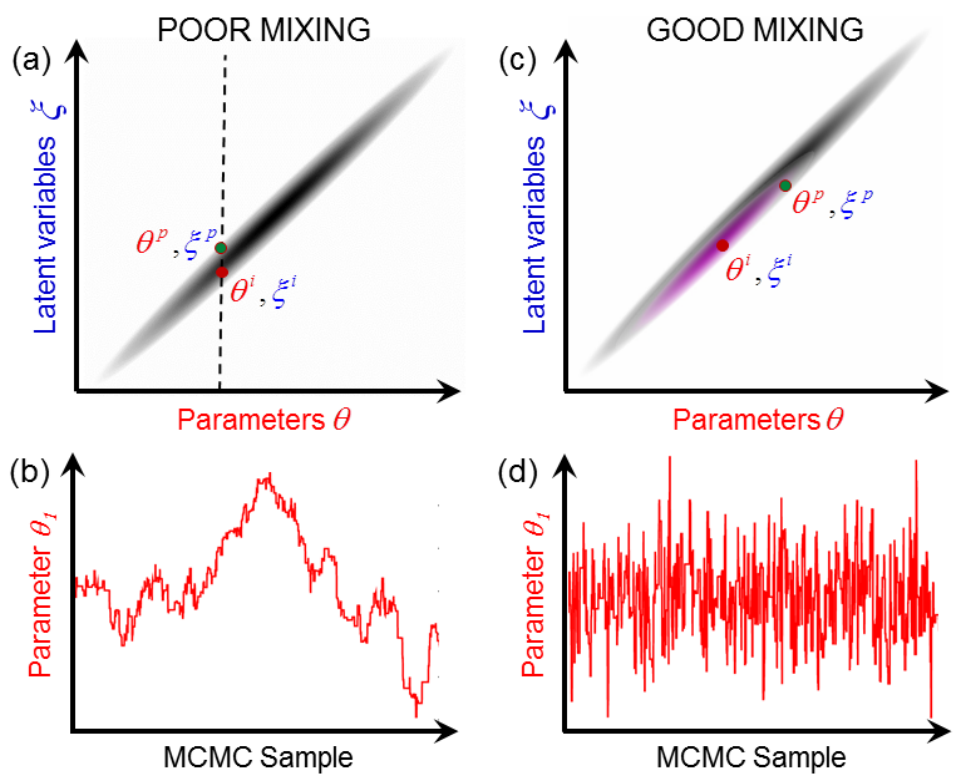

**Fig. 1: Mixing**. Illustrative example of poor/good mixing. (a) Proposals are made individually on parameters and latent variables separately (the dashed line shows the case of a latent variable being changed). The shading represents the region of high posterior probability. (b) Trace plot exhibiting poor mixing. (c) Efficient joint proposals are made using the distribution in pink (which is correlated in the same way as the posterior). (d) Trace plot exhibiting good mixing.

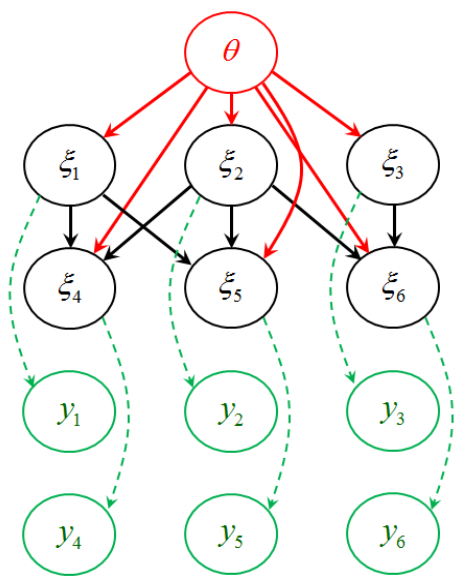

**Fig. 2: Directed acyclic graph (DAG)**. This shows a model with parameters $\theta$, latent variables $\xi$ and observations $y$. The arrows represent conditional dependencies. The model assumes that latent variables $\xi_e$ (where $e$ goes from 1 to $E=6$ in this example) are sampled from a set of univariate probability distributions $\pi(\xi_e|\xi_{e'<e},\theta)$ and $y_r$ are sampled from $\pi(y_r|\xi,\theta)$. Note, in general each observation $y_r$ can depend on an arbitrary combination of $\xi_e$ (this example shows the special case when $y_r$ depends only on $\xi_r$).

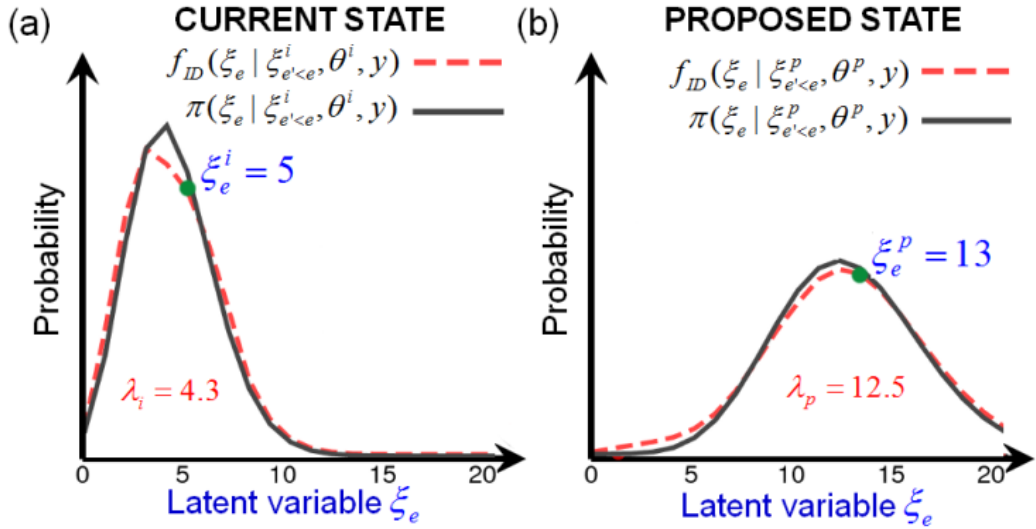

**Fig. 3: PBP updates.** Shows the true (unknown) distribution $\pi(\xi_e|\xi_{e'<e},\theta,y)$ (black lines) and importance distribution $f_{ID}(\xi_e|\xi_{e'<e},\theta,y)$ (dashed red lines) for a given latent variable $\xi_e$ using (a) the current state on the MCMC chain and (b) the proposed state. In this particular example $\xi_e$ takes non-negative integer values and the IDs are Poisson distributions.

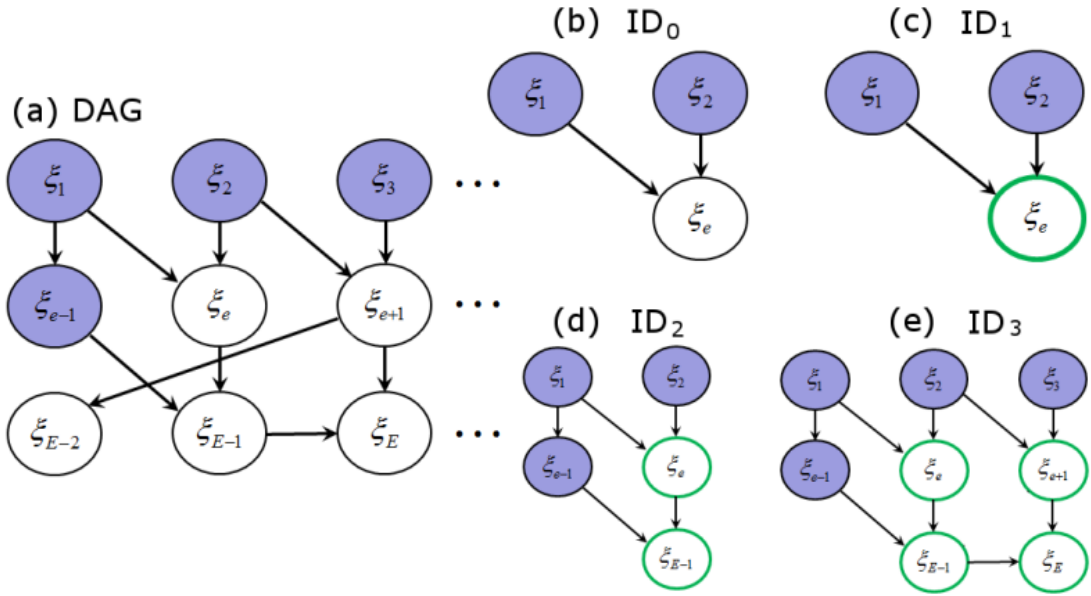

**Fig. 4: Importance distributions (ID).** (a) The model (circles containing parameters and observations have been omitted for clarity). The posterior distribution for $\xi_e$ assumes that $\xi_{e'<e}$ are known (indicated by the blue shading). (b-e) Successive approximations for the ID (described in the text). The bold, green circles indicate that observations on these latent variables have been used in calculating the corresponding ID.

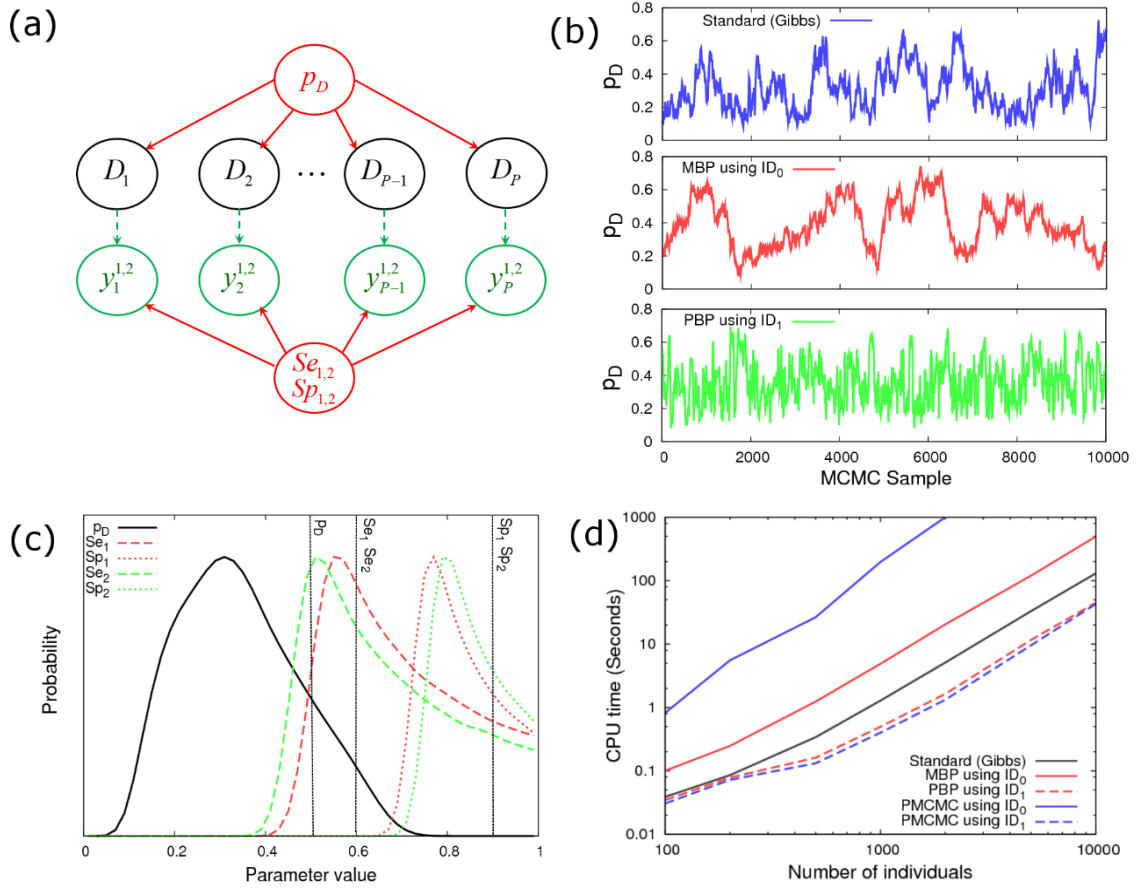

**Fig 5: Disease diagnostic test model.** (a) The DAG with probability of disease $p_D$, true disease status $D_e$ (1/0 denoting infected/uninfected), and observed disease status as measured from two diagnostic tests $y_{e1,2}$ (which have sensitivities $Se_{1,2}$ and specificities $Sp_{1,2}$). (b) Trace plots for disease probability $p_D$ as a function of MCMC sample for three different algorithms. (c) The posterior distributions for the model parameters (the vertical lines represent the true values). (d) Shows how the CPU time needed to generate 100 effective samples for $p_D$ changes as a function of the number of individuals in the population for various different approaches.

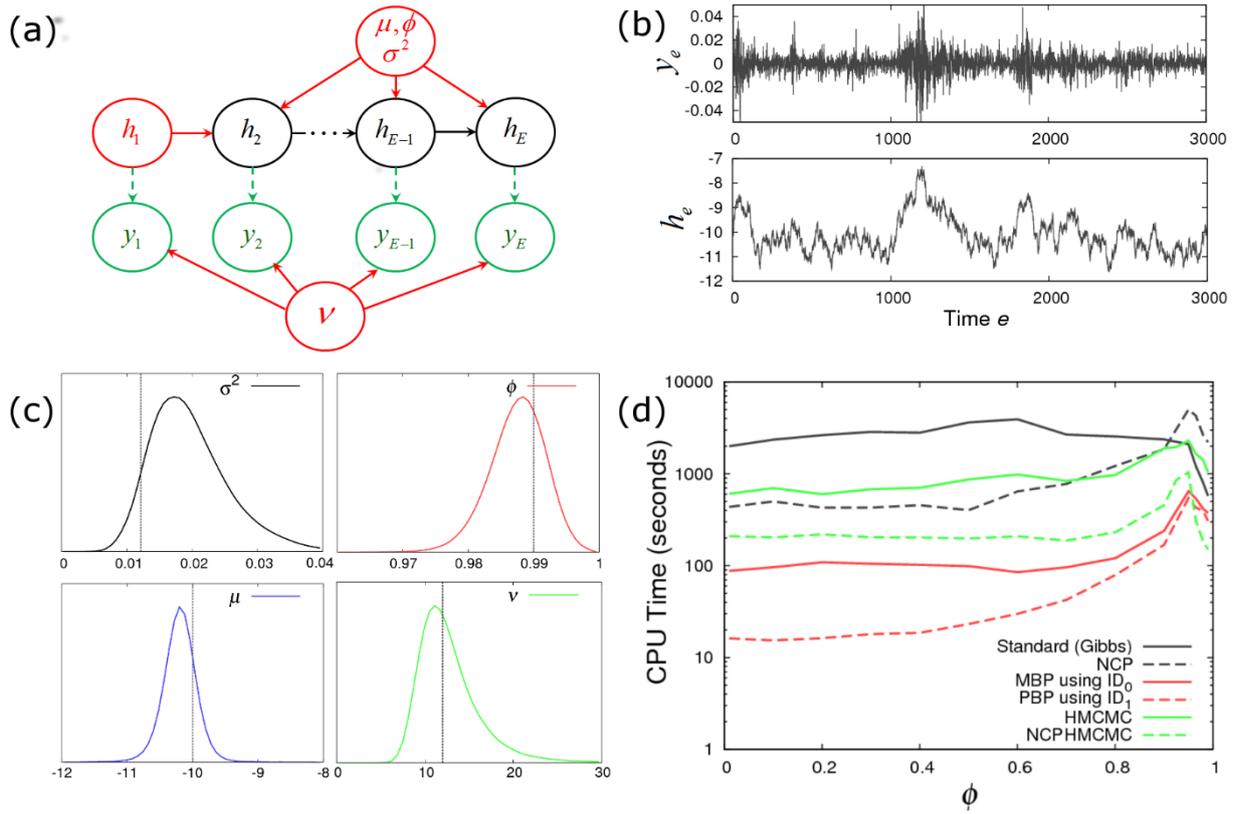

**Fig. 6: Stochastic volatility model.** (a) The DAG (see Section 5.2 for details). (b) Simulated data. (c) Posterior distributions for model parameters (the vertical lines represent the true values) based on the simulated data. (d) Shows how the CPU time needed to generate 100 effective samples of $\sigma^2$ varies as the correlation parameter $\phi$ used to simulate the data changes.

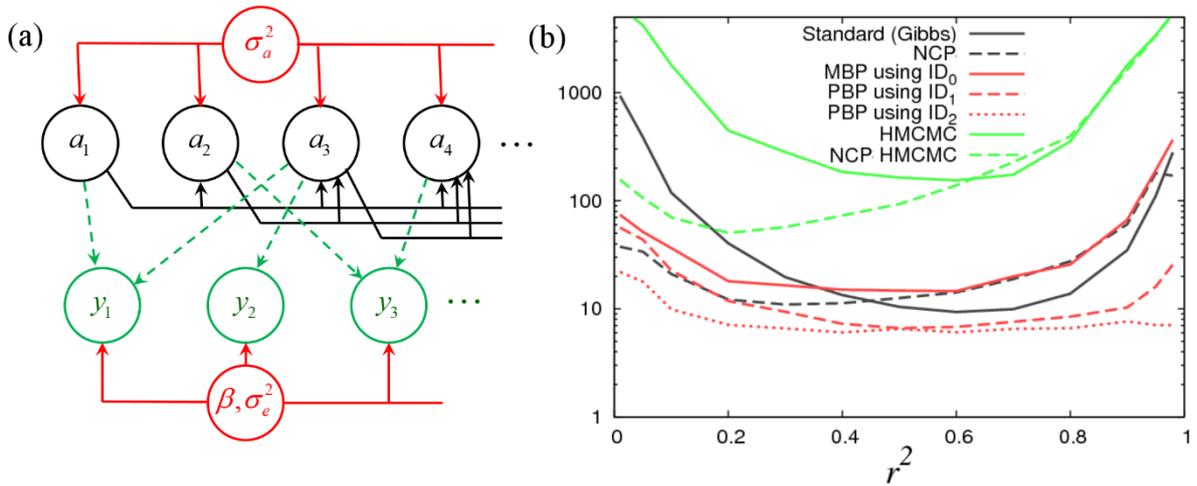

**Fig. 7: Mixed model.** This looks at a mixed model applied to a quantitative genetics application. (a) The DAG where $a$ are multivariate normally distributed with covariance matrix $\sigma_a^2 \mathbf{A}$ ($\mathbf{A}$ is the relationship matrix), vector $\beta$ are fixed effects and $y$ are trait observations with residual covariance matrix $\sigma_e^2 \mathbf{I}$ (see Section 5.3 for further details). (b) Shows how the CPU time needed to generate 100 effective samples of $r^2 = \sigma_a^2/(\sigma_a^2 + \sigma_\varepsilon^2)$ (which characterises the genetic heritability) varies as the $r^2$ used to simulate the data changes.

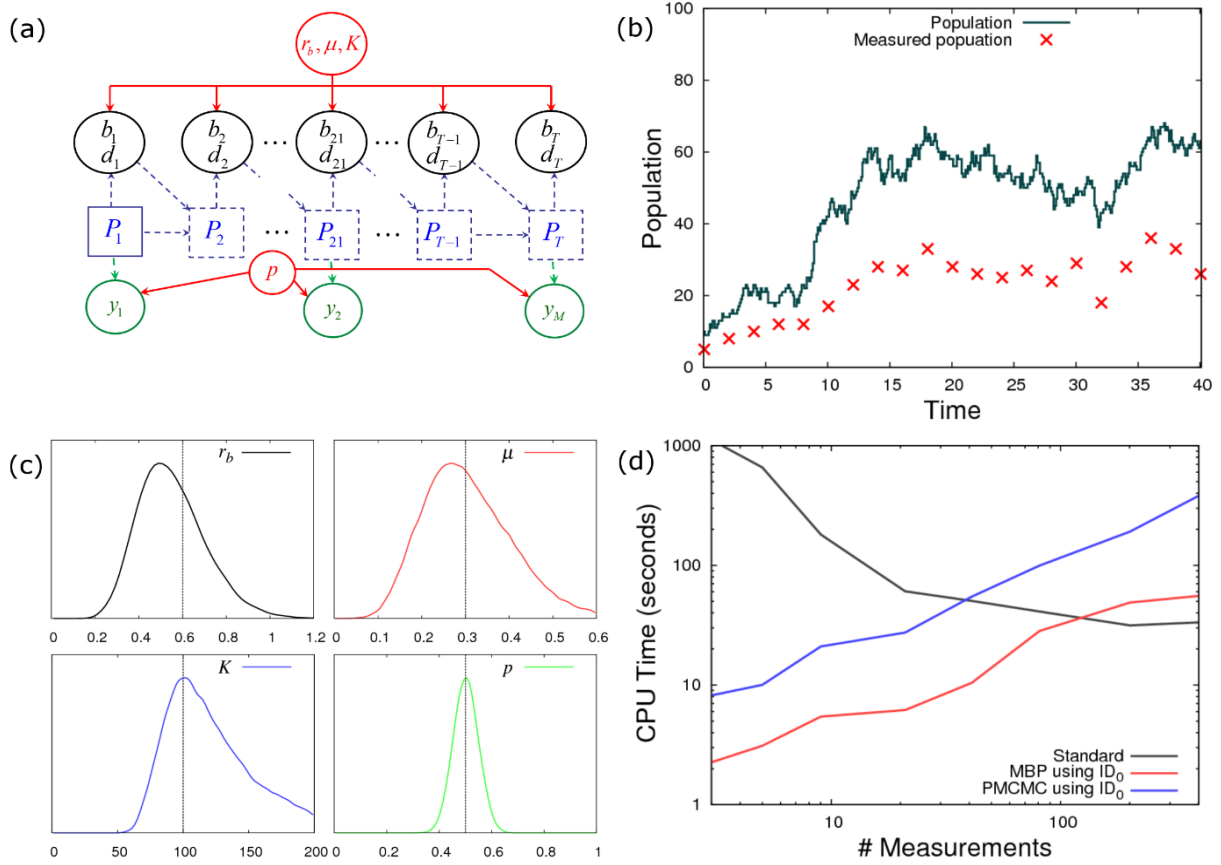

**Fig. 8: Logistic population model.** (a) The DAG, where $r_b$ is the birth rate, $\mu$ is the mortality rate, $K$ is the carrying capacity, $b_t$ and $d_t$ are the number of births and deaths at discrete time interval $t$ (which runs from 1 to $T$), $P_t$ is the population size, and $y_m$ (where $m$ runs from 1 to $M$) are the numbers of trapped individuals at various measurement points ($p$ is the trapping probability). (b) Simulated data, where $T$=401 and $\tau$=0.1 is the time step size, $r_b$=0.6, $\mu$=0.3, $K$=100, $p$=0.5 are the model parameters and $M$=21 is the number of measurements. (c) The posterior distributions for the model parameters (the vertical lines represent the true values) based on this simulated data. (d) Shows how the CPU time needed to generate 100 effective samples of $r_b$ varies as the number of equally spaced measurements $M$ changes.

## Appendix A: Importance sampling

Importance sampling (IS) is a general technique for estimating properties of a given distribution (which can't directly be sampled from) using samples generated from a different (more accessible) distribution [39, 40]. We emphasise that PBPs are *not* a type of IS, but do make use of importance distributions (IDs). Therefore a brief description is now given. The posterior in Eq.(1) can be expressed as

$$\pi(\theta, \xi | y) = \pi(\xi | \theta, y) \pi(\theta | y), \tag{A1}$$

which, using Eq.(2), may be written

$$\pi(\theta, \xi | y) = \pi(\theta | y) \prod_{e=1}^{E} \pi(\xi_e | \xi_{e'<e}, \theta, y). \tag{A2}$$

This implies that, in principle at least, posterior samples can be generated by first sampling θ from π(θ|y), then $\xi_1$ from π($\xi_1$|θ,y), then $\xi_2$ from π($\xi_2$|$\xi_1$,θ,y), and so on and so forth until a complete set of latent variables ξ is generated. Unfortunately, however, these distributions are typically intractable, so cannot be directly sampled. To overcome this difficulty importance distributions (IDs) $f_{ID}(\theta|y)$ (which, for example, could be chosen to be multivariate normal) and $f_{ID}(\xi_e|\xi_{e'<e},\theta,y)$ (a set of univariate distributions, such as normal or Poisson, for each variable *e*) are defined which *can* be sampled from. IDs are chosen to resemble the true distributions as closely as possible.

IS consists of first sampling θ from $f_{ID}(\theta|y)$ and then successively drawing $\xi_e$ from $f_{ID}(\xi_e|\xi_{e'<e},\theta,y)$ for *e*=1 to *E*. The resulting sample θ, ξ has an associated "weight"

$$w = \frac{\pi(y|\xi,\theta)\pi(\xi|\theta)\pi(\theta)}{f_{ID}(\theta|y)\prod_{e=1}^{E} f_{ID}(\xi_e|\xi_{e'<e},\theta,y)}, \tag{A3}$$

which accounts for the fact that it isn't a true posterior sample [39]. Repeating IS a sufficient number of times gives unbiased estimates for posterior quantities of interest. For example, if *i* indexes sample number then

$$\langle \theta \rangle = \sum_i \theta^i w^i \Big/ \sum_i w^i \tag{A4}$$

gives an estimate for parameter posterior means[21]. Unfortunately IS becomes highly inefficient for complex models because the vast majority of samples have negligible weight (leading to poor statistical estimates for quantities such as those in Eq.(A4)). This necessitates the use of MCMC approaches in the first place.

In summary, this appendix has identified importance distributions $f_{ID}(\xi_e|\xi_{e'<e},\theta,y)$ (which take standard functional forms such as normal, Poisson, *etc.*) that aim to account for both the model and observations by approximating π($\xi_e$|$\xi_{e'<e}$,θ,y). For example supposing that π($\xi_e$|$\xi_{e'<e}$,θ,y) has a normal-like distribution, the importance distribution would be chosen to be normal[22]

---

[21] The average of the weights themselves $\Sigma_i w^i/N$ gives an unbiased estimate of the model evidence π(y).

[22] See Appendix G for a definition of this distribution.

$f_{ID}(\xi_e|\xi_{e'<e},\theta,y)=f_{\text{norm}}(\xi_e|\mu(\xi_{e'<e},\theta,y),\sigma(\xi_{e'<e},\theta,y))$ with mean $\mu$ and standard deviation $\sigma$ functionally depend on $\xi_{e'<e}$, $\theta$ and $y$. Details on these functional dependencies are given in Section 4.

## Appendix B: Derivation of PBPs for Poisson or normal IDs

Here we explicitly demonstrate the validity of the two conditions in Eq.(6) when using the sampling procedures outlined in Table 1 for the Poisson and normal IDs.

**Poisson ID**

For a model which utilises a Poisson ID for latent variable $\xi_e$, the following probability mass functions are defined:

$$f_{ID}(\xi_e^i | \xi_{e'<e}^i, \theta^i, y) = \frac{\lambda_i^{\xi_e^i} e^{-\lambda_i}}{\xi_e^i!},$$
$$f_{ID}(\xi_e^p | \xi_{e'<e}^p, \theta^p, y) = \frac{\lambda_p^{\xi_e^p} e^{-\lambda_p}}{\xi_e^p!}, \tag{B1}$$

where $\lambda_i$ is some known function of $(\xi_{e'<e}^i, \theta^i, y)$ and $\lambda_p$ is some known function of $(\xi_{e'<e}^p, \theta^p, y)$.

Suppose, arbitrarily, that the expected number of occurrences in the proposed state $\lambda_p$ is greater than the expected number in the initial state $\lambda_i$. Table 1 shows that the actual number in the proposed state $\xi_e^p$ is calculated using

$$\xi_e^p = \xi_e^i + X, \tag{B2}$$

where $\xi_e^i$ is the number in the initial state and $X$ is sampled from a Poisson distribution with average number given by the difference between $\lambda_p$ and $\lambda_i$:

$$X \sim \text{Pois}(\lambda_p - \lambda_i). \tag{B3}$$

The probability of this proposal is given by

$$g(\xi_e^p) = \frac{(\lambda_p - \lambda_i)^{\xi_e^p - \xi_e^i} e^{-(\lambda_p - \lambda_i)}}{(\xi_e^p - \xi_e^i)!}. \tag{B4}$$

Interchanging $i$ and $p$ in Table 1 shows that the reverse transition is taken from a binomial probability distribution

$$\xi_e^i \sim B(\xi_e^p, \tfrac{\lambda_i}{\lambda_p}), \tag{B5}$$

with proposal probability

$$g(\xi_e^i) = \frac{\xi_e^p!}{\xi_e^i!(\xi_e^p - \xi_e^i)!} \left(\tfrac{\lambda_i}{\lambda_p}\right)^{\xi_e^i} \left(1 - \tfrac{\lambda_i}{\lambda_p}\right)^{\xi_e^p - \xi_e^i}. \tag{B6}$$

Combining the results from Eqs.(B4) and (B6) gives

$$\frac{g(\xi_e^i)}{g(\xi_e^p)} = \frac{\xi_e^p!}{\xi_e^i!(\xi_e^p - \xi_e^i)!}\left(\frac{\lambda_i}{\lambda_p}\right)^{\xi_e^i}\left(1 - \frac{\lambda_i}{\lambda_p}\right)^{\xi_e^p - \xi_e^i} \times \left(\frac{(\lambda_p - \lambda_i)^{\xi_e^p - \xi_e^i} e^{-(\lambda_p - \lambda_i)}}{(\xi_e^p - \xi_e^i)!}\right)^{-1}$$

$$= \frac{\lambda_p^{\xi_e^i} e^{-\lambda_i}}{\xi_e^i!} \times \left(\frac{\lambda_p^{\xi_e^p} e^{-\lambda_p}}{\xi_e^p!}\right)^{-1}.$$

(B7)

Using this, along with Eq.(B1), finally gives

$$\frac{f_{ID}(\xi_e^p | \xi_{e'<e}^p, \theta^p, y)}{f_{ID}(\xi_e^i | \xi_{e'<e}^i, \theta^i, y)} \frac{g(\xi_e^i)}{g(\xi_e^p)} = \frac{\lambda_p^{\xi_e^p} e^{-\lambda_p}}{\xi_e^p!}\left(\frac{\lambda_i^{\xi_e^i} e^{-\lambda_i}}{\xi_e^i!}\right)^{-1} \frac{\lambda_p^{\xi_e^i} e^{-\lambda_i}}{\xi_e^i!}\left(\frac{\lambda_p^{\xi_e^p} e^{-\lambda_p}}{\xi_e^p!}\right)^{-1} = 1,$$

(B8)

which shows that condition 1 in Eq.(6) is, indeed, satisfied. Condition 2 is satisfied because $X=0$ when $\lambda_p = \lambda_i$ in Eq.(B3), and so by definition $\xi^p = \xi^i$ in Eq.(B2) (similarly $\xi^i = \xi^p$ in Eq.(B5)).

**Normal ID**

Here we consider the case of a normal ID for latent variable $\xi_e$:

$$f_{ID}(\xi_e^i | \xi_{e'<e}^i, \theta^i, y) = \frac{1}{\sqrt{2\pi}\sigma_i} e^{-\frac{(\xi_e^i - \mu_i)^2}{2\sigma_i^2}},$$

$$f_{ID}(\xi_e^p | \xi_{e'<e}^p, \theta^p, y) = \frac{1}{\sqrt{2\pi}\sigma_p} e^{-\frac{(\xi_e^p - \mu_p)^2}{2\sigma_p^2}}.$$

(B9)

Suppose, arbitrarily, that the standard deviation in the proposed state $\sigma_p$ is smaller than in the initial state $\sigma_i$. Table 1 shows that the proposal for $\xi_e^p$ is given by

$$\xi_e^p \sim N\left(\mu_p + \alpha \frac{\sigma_p^2}{\sigma_i^2}(\xi_e^i - \mu_i), \kappa \frac{\sigma_p^2}{\sigma_i^2}(\sigma_i^2 - \sigma_p^2)\right),$$

(B10)

where $\alpha^2 = \kappa + (1-\kappa)\frac{\sigma_i^2}{\sigma_p^2}$ and $\kappa$ is a tuneable constant. The probability density function for generating this final state is given by

$$g(\xi_e^p) = \frac{1}{\sqrt{2\pi\kappa\frac{\sigma_p^2}{\sigma_i^2}(\sigma_i^2 - \sigma_p^2)}} e^{-\frac{\left(\xi_e^p - \left(\mu_p + \alpha\frac{\sigma_p^2}{\sigma_i^2}(\xi_e^i - \mu_i)\right)\right)^2}{2\kappa\frac{\sigma_p^2}{\sigma_i^2}(\sigma_i^2 - \sigma_p^2)}}.$$

(B11)

We now consider the reverse transition. Since *p* and *i* are now switched in Table 1, the corresponding proposal probability density function is given by

$$g(\xi_e^i) = \frac{1}{\sqrt{2\pi\kappa(\sigma_i^2 - \sigma_p^2)}} e^{-\frac{\left(\xi_e^i - \left(\mu_i + \alpha(\xi_e^p - \mu_p)\right)\right)^2}{2\kappa(\sigma_i^2 - \sigma_p^2)}}. \quad (B12)$$

The ratio between Eqs.(B12) and (B11) is given by

$$\begin{aligned}
\frac{g(\xi_e^i)}{g(\xi_e^p)} &= \sqrt{\frac{\sigma_p^2}{\sigma_i^2}} e^{-\frac{1}{2\kappa(\sigma_i^2 - \sigma_p^2)}\left[\left((\xi_e^i - \mu_i) - \alpha(\xi_e^p - \mu_p)\right)^2 - \frac{\sigma_i^2}{\sigma_p^2}\left((\xi_e^p - \mu_p) - \alpha\frac{\sigma_p^2}{\sigma_i^2}(\xi_e^i - \mu_i)\right)^2\right]} \\
&= \sqrt{\frac{\sigma_p^2}{\sigma_i^2}} e^{-\frac{1}{2\kappa(\sigma_i^2 - \sigma_p^2)}\left[(\xi_e^i - \mu_i)^2 + \alpha^2(\xi_e^p - \mu_p)^2 - 2\alpha(\xi_e^i - \mu_i)(\xi_e^p - \mu_p) - \left(\frac{\sigma_i^2}{\sigma_p^2}(\xi_e^p - \mu_p)^2 + \alpha^2\frac{\sigma_p^2}{\sigma_i^2}(\xi_e^i - \mu_i)^2 - 2\alpha(\xi_e^p - \mu_p)(\xi_e^i - \mu_i)\right)\right]} \\
&= \sqrt{\frac{\sigma_p^2}{\sigma_i^2}} e^{-\frac{1}{2\kappa(\sigma_i^2 - \sigma_p^2)}\left[\left(1 - \alpha^2\frac{\sigma_p^2}{\sigma_i^2}\right)(\xi_e^i - \mu_i)^2 + \left(\alpha^2 - \frac{\sigma_i^2}{\sigma_p^2}\right)(\xi_e^p - \mu_p)^2\right]} \\
&= \sqrt{\frac{\sigma_p^2}{\sigma_i^2}} e^{-\frac{\sigma_i^2 - \alpha^2\sigma_p^2}{2\kappa(\sigma_i^2 - \sigma_p^2)}\left[\frac{1}{\sigma_i^2}(\xi_e^i - \mu_i)^2 - \frac{1}{\sigma_p^2}(\xi_e^p - \mu_p)^2\right]}.
\end{aligned} \quad (B13)$$

The definition $\alpha^2 = \kappa + (1-\kappa)\frac{\sigma_i^2}{\sigma_p^2}$ can be rearrange to $\kappa(\sigma_i^2 - \sigma_p^2) = \sigma_i^2 - \alpha^2\sigma_p^2$, leading to

$$\begin{aligned}
\frac{g(\xi_e^i)}{g(\xi_e^p)} &= \sqrt{\frac{\sigma_p^2}{\sigma_i^2}} e^{-\frac{\sigma_i^2 - \alpha^2\sigma_p^2}{2\kappa(\sigma_i^2 - \sigma_p^2)}\left[\frac{1}{\sigma_i^2}(\xi_e^i - \mu_i)^2 - \frac{1}{\sigma_p^2}(\xi_e^p - \mu_p)^2\right]} \\
&= \frac{1}{\sqrt{2\pi\sigma_i^2}}\left(e^{-\frac{1}{2\sigma_i^2}(\xi_e^i - \mu_i)^2}\right)\left[\frac{1}{\sqrt{2\pi\sigma_p^2}}\left(e^{-\frac{1}{2\sigma_p^2}(\xi_e^p - \mu_p)^2}\right)\right]^{-1}.
\end{aligned} \quad (B14)$$

This, together with the expressions in Eq.(B9), shows that condition 1 in Eq.(6) is, indeed, satisfied. Furthermore, condition 2 is satisfied by noting that the variance in the proposal in Eq.(B10) goes to zero when $\sigma_p^2 = \sigma_i^2$.

The validity of all the sampling procedures in Table 1 can be verified by following essentially the same procedure as above to show that the two conditions in Eq.(6) are satisfied.

## Appendix C: Adaptation period

Posterior-based proposals contain two quantities in Eq.(8) that need to be established: a numerical approximation to the covariance matrix of the posterior $\Sigma^\theta$, and a jumping size $j$. Motivated by adaptive MCMC [41, 42], these are calculated during an "adaptation" period, which also acts as the required burn-in phase (*i.e.* this ensures that the first sample drawn after the adaptation period is representative of a random draw from the posterior). In this study $I^{ad}=10^4$ iterations are used for adaptation, and subsequently $\Sigma^\theta$ and $j$ are fixed[23].

---

[23] This ensures that detailed balance is strictly enforced when MCMC samples are actually taken.

**Covariance matrix $\Sigma_\theta$**: During the adaptation period the number of MCMC iterations *i* changes from 1 to $I^{ad}$. An approximation to the posterior covariance matrix $\Sigma^\theta$ is calculated every *100* iterations based on samples from *i/2* to *i*:

$$\Sigma^\theta_{mn} = \frac{1}{\frac{1}{2}i - 1} \sum_{i'=\frac{i}{2}}^{i} (\theta^{i'}_m - \overline{\theta}_m)(\theta^{i'}_n - \overline{\theta}_n), \quad \overline{\theta}_m = \frac{1}{\frac{1}{2}i} \sum_{i'=\frac{i}{2}}^{i} \theta^{i'}_m. \tag{C1}$$

This is effectively equivalent to using a dynamically changing burn-in period set at half the current iteration number. As the adaptation period progresses the estimated covariance matrix $\Sigma^\theta$ becomes a better and better approximation, which helps to improve the efficiency of the algorithm. For the first 100 samples $\Sigma^\theta$ is set to a diagonal matrix with elements chosen to be sufficiently small to ensure a good initial acceptance rate.

**Jumping size *j***: This determines the acceptance rate for PBPs in Eq.(9). If *j* is too large very few proposals are accepted, and if too small mixing is slow. A robust heuristic method for optimising *j* is as follows. Initially *j* is set to a small quantity. Each time a PBP is accepted, *j* is updated according to

$$j^{new} = j \times 1.02, \tag{C2}$$

and when rejected

$$j^{new} = j \times 0.99. \tag{C3}$$

These numerical factors are chosen for two reasons: Firstly, the two updates in Eqs.(C2) and (C3) balance each other out when acceptance occurs around 33% of the time (which from Appendix H was found to be approximately optimal), leading to a steady state solution for *j*. Secondly, they are sufficiently close to 1 to prevent large fluctuations in *j*, but sufficiently far to allow for the steady state solution to be found during the adaptation period.

## Appendix D: Derivation of acceptance probability

We derive the expression in Eq.(9). Based on Eq.(1), the MH acceptance probability is given by

$$P_{MH} = \min\left\{1, \frac{\pi(y|\xi^p,\theta^p)\pi(\xi^p|\theta^p)\pi(\theta^p)}{\pi(y|\xi^i,\theta^i)\pi(\xi^i|\theta^i)\pi(\theta^i)} \frac{g^{p\to i}}{g^{i\to p}}\right\}, \tag{D1}$$

where $g^{i\to p}$ represents the proposal probability density for generating $\theta^p$ and $\xi^p$ given the current state $\theta^i$ and $\xi^i$, and $g^{p\to i}$ represents the corresponding quantity in the opposite direction[24]. Following steps 1 and 2 in the PBP algorithm from Section 3.4, the overall proposal probability can be expressed as

$$g^{i\to p} = f_{MVN}(\theta^p - \theta^i, j^2\Sigma^\theta) \prod_{e=1}^{E} g(\xi^p_e). \tag{D2}$$

Substituting this, along with the reverse transition, into Eq.(D1) (and noting that the MVN distributions are symmetric[25]), gives

---

[24] In other words, starting at $\theta^p$ and $\xi^p$ and proposing the state $\theta^i$ and $\xi^i$.
[25] The probability of jumping from $\theta^i$ to $\theta^p$ is the same as from $\theta^p$ to $\theta^i$.

$$P_{MH} = \min\left\{1, \frac{\pi(y|\xi^p,\theta^p)\pi(\xi^p|\theta^p)\pi(\theta^p)}{\pi(y|\xi^i,\theta^i)\pi(\xi^i|\theta^i)\pi(\theta^i)} \prod_{e=1}^{E} \frac{g(\xi_e^p)}{g(\xi_e^i)}\right\}. \tag{D3}$$

Substituting condition 1 from Eq.(6) into this expression leads to the final result in Eq.(9).

## Appendix E: Further insights into PBPs

Here we provide some additional notes on the PBP algorithm in Section 3.4:

**Step 1:** Strong posterior correlations can exist not only between model parameters and latent variables (as demonstrated in Fig. 1(a)), but also between different model parameters themselves (*i.e.* after marginalisation over latent variables). Equation (8) helps to mitigate against these, helping to further facilitating mixing. Other possibilities for proposals in parameter space are discussed in Appendix F (along with complications such as what to do when parameters are discrete rather than continuous), and sampling from MVNs using Cholesky decomposition [43] is described in Appendix G.

**Step 2:** This step makes use of IDs $f_{ID}(\xi_e|\xi_{e'<e},\theta,y)$, which approximate the univariate distributions $\pi(\xi_e|\xi_{e'<e},\theta,y)$ for $e=1,…,E$ (with functional form chosen to provide good approximation to the model under study). Each functional form for the IDs is associated with a different (posterior based) proposal distributions $g(\xi_e^p)$ satisfying Eq.(6) (see Table 1). IDs are characterised by one or two parameters (*e.g.* an expected event number in the Poisson case and a mean and variance in the normal case) and for $e>1$ these functionally depend on latent variables with lower index (as determined by the DAG structure of the underlying the model) and the model parameters and data.

**Step 3:** Two limiting cases in Eq.(9) are of particular importance. Firstly, as $f_{ID}(\xi_e|\xi_{e'<e},\theta,y)$ becomes more and more representative of $\pi(\xi_e|\xi_{e'<e},\theta,y)$, the MH probability reduces to

$$P_{MH} = \min\left\{1, \frac{\pi(\theta^p|y)}{\pi(\theta^i|y)}\right\}. \tag{E1}$$

The originally high dimensional problem (containing latent variables ξ and parameters θ) is reduced to a much lower dimensional problem (containing just parameters θ), helping explain why mixing is potentially so much faster.

Secondly, as the jumping size in parameter space gets smaller (as determined by *j* in Eq.(8)) so $P_{MH}$→1. This is of particular importance because it means that even if the IDs provide a relatively poor approximation to $\pi(\xi_e|\xi_{e'<e},\theta,y)$, provided *j* is made sufficiently small a substantial fraction of proposals will always be accepted. In practice the jumping size *j* is optimally tuned to ensure acceptance around 33% of the time (see Appendix C for details).

## Appendix F: Other possibilities for proposals in parameter space

Three issues are discussed in relation to proposals in parameter space:

### 1) Optimisation

The proposal distribution in parameter space introduced in Eq.(8) has the advantage of being simple and robust against highly correlated model parameters. Generally speaking, however, it may not represent the optimum choice. For example, if two variables A and B are largely uncorrelated in the posterior it may actually be computationally faster to consider proposals to A and B separately. This is especially true in cases when proposing changes to A is much faster (*e.g.* fixed effects in mixed models) than B (*e.g.* random effects).

In the most general case, a combination of the following two types of proposal can be used in Eq.(8):

**Single parameters changes** – A single parameter *k* is selected from $\theta$. $\theta_k^p$ is then sampled from a simple normal distribution centred at $\theta_k^i$:

$$\theta_k^p \sim N(\theta_k^i, \sigma_k^2), \quad \theta_{l \neq k}^p = \theta_l^i. \tag{F1}$$

**Multiple parameter changes** – $\chi$ represent a subset of the parameters in $\theta$, and $\theta_\chi^p$ is sampled from a multivariate normal distribution centred at $\theta_\chi^i$:

$$\theta_\chi^p \sim N(\theta_\chi^i, j_\chi^2 \Sigma^\chi), \quad \theta_{l \notin \chi}^p = \theta_l^i. \tag{F2}$$

Providing an automated way to determine the optimum choice for proposals in parameter space for a given model will be the subject of future research.

### 2) Parameters not continuous

In some cases model parameters are discrete, *e.g.* for an epidemiological model the initial population numbers in various compartments might be included in θ. If these discrete variables are approximately normally distributed (as they would be if the population sizes are reasonably large), then we could again draw a vector *v* from a MVN distribution as before

$$v \sim N(\theta^i, j^2 \Sigma^\theta), \tag{F3}$$

but this time round those discrete model variables to the nearest integer

$$\theta_k^p = \begin{cases} \text{round}(v_j), & \text{if } \theta_j \text{ discrete} \\ v_j, & \text{if } \theta_j \text{ continuous} \end{cases}. \tag{F4}$$

In cases in which model parameters are not expected to be approximately normally distributed they can simply be updated individually.

### 3) Restrictions

For some of the functional forms in Table 1, only one of the characteristic parameters can be updated at a time. For example, for the beta distribution only α can be changed whilst fixing β, or *vice versa*. Consequently, proposals to both α and β cannot be performed simultaneously.

## Appendix G: Normal distributions

The probability density function for drawing a value $x$ from a normal distribution with mean $\mu$ and variance $\sigma^2$ is given by

$$f_{norm}(x\mid\mu,\sigma^2) = \frac{1}{\sqrt{2\pi\sigma^2}} e^{-\frac{(x-\mu)^2}{2\sigma^2}}. \qquad (G1)$$

The equivalent multivariate normal distribution is

$$f_{MVN}(x\mid\boldsymbol{\mu},\boldsymbol{\Sigma}) = \frac{1}{\sqrt{(2\pi)^d|\boldsymbol{\Sigma}|}} e^{-\frac{1}{2}(x-\mu)^T \Sigma^{-1} (x-\mu)}, \qquad (G2)$$

where $d$ is the number of dimensions and $\boldsymbol{\Sigma}$ is the covariance matrix that captures the variance of individual elements (*e.g.* parameters) as well as covariance between them.

Cholesky decomposition provides a standard way to draw samples from a multivariate normal distribution [43]. Provided $\boldsymbol{\Sigma}$ is positive-definite it can be written as $\boldsymbol{\Sigma}=\boldsymbol{B}\boldsymbol{B}^T$, where $\boldsymbol{B}$ is a lower triangular matrix. Samples from the multivariate normal are then generated using

$$\boldsymbol{x} = \boldsymbol{\mu} + \boldsymbol{B}\boldsymbol{z}, \qquad (G3)$$

where $\boldsymbol{z}$ is a vector of normally distributed independent samples with mean zero and unit variance.

## Appendix H: Optimisation of the PBP MCMC algorithm

Figure H illustrates how the PBP algorithm is typically optimised. These results are based on the mixed model applied to quantitative genetics introduced in Section 5.3, but the general findings are found to be largely independent of model type. Optimisation can be considered from three points of view:

Firstly, Fig. H(a) shows the CPU time to generate 100 effective samples of $r^2$ from the posterior as a function of the number of PBPs between each Gibbs update of the latent variables. We find that this particular curve has a minimum at $U=4$ updates, although computation speed is found to not be particularly sensitive to its exact value.

Secondly, Fig. H(b) shows variation in CPU time as a function of the tuneable constant $\kappa$ (this is used in Table 1 in cases in which the ID is normally distributed). Again, performance is largely the same

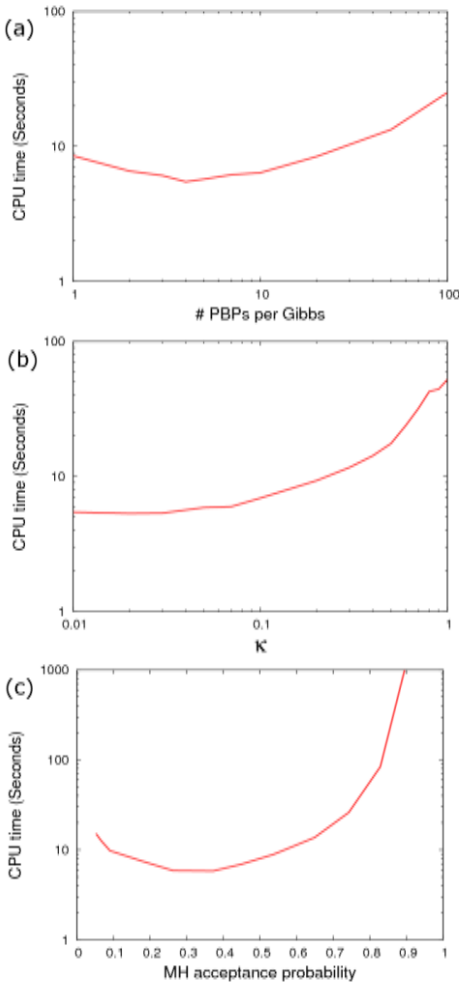

**Figure H:** Optimisation of key parameters.

provided $\kappa$ is smaller than around 0.1. For this study $\kappa$=0.03 was selected.

Finally, Fig. H(c) shows that the algorithm is optimised when the MH acceptance probability is around 33%. This is implemented using the methods outlined in Appendix C.

## Appendix I: Non-centred parameterisations

It has long been recognized that the parametrization of hierarchical models can be crucial for MCMC performance [26]. A so-called "centred" parameterisation (CP) is the default option given by the specification of the model in terms of parameters θ which determine the distribution of latent variables ξ. On the other hand "non-centred" parameterisation (NCP) refers to the case in which a new set of latent variables ξ' are defined so as to be distributed conditionally independently of θ, and the original latent variables are functionally dependant on these, *i.e.* ξ=*h*(ξ',θ,*y*).

To give a simple example, suppose each latent variable is distributed normally $\xi_e \sim N(\mu, \sigma^2)$ with mean $\mu$ and variance $\sigma^2$ being model parameters θ. This can be reparametrized by setting $\xi'_e \sim N(0,1)$, with the functional dependency *h* being given by $\xi_e = \mu + \sigma \xi'_e$.

## Appendix J: Hamiltonian MCMC

The reason standard approaches (involving small local changes) are slow is because they behave diffusively. One proposal might move a parameter in one direction, but the next might move it back again to near where it started. Such random walk behaviour often leads to slow progress from one side of the posterior to the other, which is especially true for high dimensional problems. The idea behind HMCMC is to make large jumps in parameter space to overcome this diffusive behaviour.

HMCMC [3, 12] makes no distinction between model parameters and latent variables, and so subsequently we refer to the combination *x*=(θ,ξ) to represent a vector giving the overall parameters in the model. The intuition behind HMCMC comes from physics. We first define *U(x)*=-log(π(*y*|*x*)π(*x*)) as the negative log of the posterior probability, where *U(x)* maps out a potential energy landscape, and consider a particle moving in this space. The particle has both a position vector *x* and a momentum vector ***p***. Just as a ball on a hill runs down and accelerates, so a particle with high potential *U* gets pushed towards lower *U*, at the same time increasing its kinetic energy. An important principle in physics is the conservation of energy. Here we defined the total energy of the system by the Hamiltonian

$$H(\boldsymbol{x}, \boldsymbol{p}) = \tfrac{1}{2} \boldsymbol{p}^T \boldsymbol{M}^{-1} \boldsymbol{p} + U(\boldsymbol{x}), \tag{J1}$$

where the first term represents the kinetic energy (***M*** is the mass matrix, as identified below) and the second term represents the potential energy.

The following algorithm outlines the procedure for a single HMCMC update.

---

*HMCMC algorithm*

**Step 1: Sample momentum** – The initial momentum vector at time *t*=0 is sampled according to

$$p(0) \sim N(0, M), \quad (J2)$$

and *x*(0) is set to the current parameter set $x^i$ on the MCMC chain.

**Step 2: Integration of trajectory** – The following leapfrog algorithm is iterated *L* times:

$$p(t+\tfrac{\varepsilon}{2}) = p(t) - \frac{\varepsilon}{2}\nabla_x U\big|_t,$$

$$x(t+\varepsilon) = x(t) + \varepsilon M^{-1} p(t+\tfrac{\varepsilon}{2}), \quad (J3)$$

$$p(t+\varepsilon) = p(t+\tfrac{\varepsilon}{2}) - \frac{\varepsilon}{2}\nabla_x U\big|_{t+\varepsilon},$$

where ε is the integration step size and $\nabla_x U|_t$ is the gradient in the potential energy evaluated at time *t* (note, this vector points uphill)**.** This procedure represents a numerical approximation to Hamilton's equations.

**Step 3: Accept or reject** – The final proposed state $x^p = x(L\varepsilon)$ is accepted or rejected with MH probability functionally dependent on the difference in the Hamiltonian between the initial and final states:

$$P_{MH} = \min\left\{1, e^{H(x(0),p(0)) - H(x(L\varepsilon),p(L\varepsilon))}\right\}. \quad (J4)$$

---

Note because the Hamiltonian is conserved, Eq.(J4) is expected to be near to one. The reason it isn't exactly one is because the continuous integral is numerically approximated by the discrete Leapfrog method (consequently $P_{MH}$→1 as ε→0, but if ε is large $P_{MH}$ can become very small if a long trajectory is integrated over).

## Optimisation

HMCMC is most efficient when the inverse of the mass matrix $M^{-1}$ is given by a numerical approximation to the covariance matrix for the posterior distribution π(*x*|*y*). However in high dimensional situations this is usually too computationally demanding to calculate (*e.g.* matrix inversion takes of order $D^3$ operations, where *D* is the number of dimensions). Instead two possibilities are commonly implemented: either set *M* to the identity matrix or set it to be diagonal with elements given by the inverse of the posterior parameter variances. Here we chose the latter option, as this was found to improve mixing times compared to the former. The algorithm above contains two tuneable parameters: step size ε and number of steps *L*. Optimising the step size ε is relatively easy, as it can be selected to achieve a certain average acceptance rate. If ε is very small then the acceptance probability will be almost one but computation will be slowed down because more and more intermediary steps will be needed for a certain integration length *Lε*. On the other hand if ε is too large, most proposals get rejected. The optimal value (under some strong assumptions) has been shown to be approximately 0.65 [13], which is used here (although efficiency was not found to be very sensitive to this precise value).

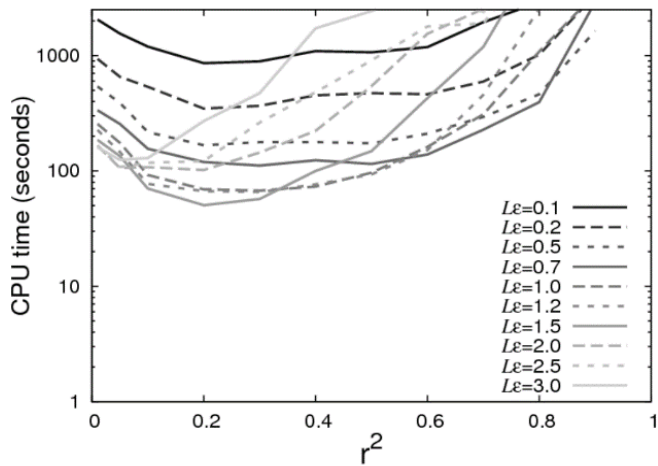

**Figure J: Optimisation of HMCMC.** These results are applicable to the mixed model in Section 5.3 and shows how the CPU time needed to generate 100 effective samples of $r^2$ (which characterises the genetic heritability) varies as the $r^2$ used to simulate the data changes. Each of the curves corresponds to running NCP HMCMC using different fixed integration lengths *Lε* (ε is adaptively tuned to give an acceptance probability of 0.65). The curve defined by the lowest points in this diagram represents the optimised NCP HMCMC results, as shown by the green dashed line in Fig.7(b).

Optimising the number of steps for each update *L*, however, is difficult and efficiency is found to critically depend on this value. Automated methods such as No U-Turn Samplers (NUTs) [13] have been developed, but these are challenging to implement. This paper takes a brute force approach to find the optimal HCMC implementation used to compare with other methods. For each set of simulated data inference is carried out using a large number of different values of integral length *Lε*, and the most efficient of these is used to construct the HMCMC curves in Figs. 6(d) and 7(b). An example of this process is shown in Fig. J, which demonstrates how the NCP HMCMC results in Fig. 7(b) were generated. Note, under realistic models the optimal results from NUTs are found to have very similar computational efficiency to HMCMC tuned in this fashion [13].

## Appendix K: Particle MCMC

The idea behind PMCMC is that random walk MCMC can be run on the basis of an unbiased approximation to $\pi(y|\theta)$. Because the dimensionality of θ is typically much less than ξ, this algorithm is expected to mix at a much faster rate than standard MCMC. The drawback of this approach, however, is that obtaining a sufficiently accurate estimate $\hat{\pi}(y|\theta)$ for $\pi(y|\theta)$ can be computationally demanding.

The algorithm below describes the implementation used in this paper [44][26] :

---

[26] Note, this method is known as the "particle marginal Metropolis–Hastings" (PMMH) sampler in this reference. The proposals in parameter space in Eq.(K1) are chosen to be consistent with Eq.(8) to allow for fair comparison between methods.

*PMCMC algorithm*

**Step 1: Generate θ$^p$** – This is the same as for PBPs. A proposed set of parameter values is drawn from a multivariate normal (MVN) distribution centred on the current set of parameters in the chain θ$^i$

$$\theta^p \sim N(\theta^i, j^2 \Sigma^\theta), \tag{K1}$$

where $\Sigma^\theta$ is a numerical approximation to the covariance matrix for $\pi(\theta|y)$ and $j$ is a tuneable jumping parameter (estimation of $\Sigma^\theta$ and optimisation of $j$ are achieved during an initial "adaptation" period, as explained in Appendix C).

**Step 2: Generate unbiased estimate $\hat{\pi}(y|\theta^p)$** – We take each latent variable $\xi_e$ in turn (starting from $e=1$ up to $e=E$) and consider $Z$ particles. The weights for these particles are initially set to $w_z=1$. For each particle $z$ we sample from an importance distribution

$$\xi_e^z \sim f_{ID}(\xi_e^z | \xi_{e'<e}^z, \theta^p, y). \tag{K2}$$

In the simplest case this will be ID$_0$, which is equivalent to simulating from the model, but as with PBPs, greater efficiency can be achieved by using higher order importance distributions. Here we imagine the case in which an observation $y_e$ is made on each latent variable with observation probability $\pi(y_e|\xi_e,\theta)$. The weight for each particle $w_z$ is then multiplied by

$$\pi(y_e | \xi_e^z, \theta^p) \frac{\pi(\xi_e^z | \xi_{e'<e}^z, \theta^p)}{f_{ID}(\xi_e^z | \xi_{e'<e}^z, \theta^p, y)}. \tag{K3}$$

After scanning through all latent variables, an unbiased estimator can be generated by

$$\pi(y | \theta^p) = \tfrac{1}{Z} \sum_{z=1}^{Z} w_z, \tag{K4}$$

which is essentially a standard implementation of importance sampling (see Appendix A). This estimator, however, turns out to usually be very computationally wasteful because most of the particles have almost zero weight and contribute very little to the sum. A key innovation in PMCMC is the introduction of "bootstrap" steps. At various points when scanning from $e=1$ to $E$, a new set of particles is sampled from the existing set with probability proportional to the particle weights. This new set then has its weights returned back to $w_z=1$ and the process is continued. Now

$$\pi(y | \theta^p) = \prod_b \left( \tfrac{1}{Z} \sum_{z=1}^{Z} w_z^b \right) \tag{K5}$$

is an unbiased estimator, where the product goes over bootstrap steps and $w_z^b$ are the weights of particles immediately prior to the bootstrap being performed.

**Step 3: Accept or reject** – The final proposed state θ$^p$ is accepted or rejected with MH probability

$$P_{MH} = \min \left\{ 1, \frac{\pi(y|\theta^p)\pi(\theta^p)}{\pi(y|\theta^i)\pi(\theta^i)} \right\}. \tag{K6}$$

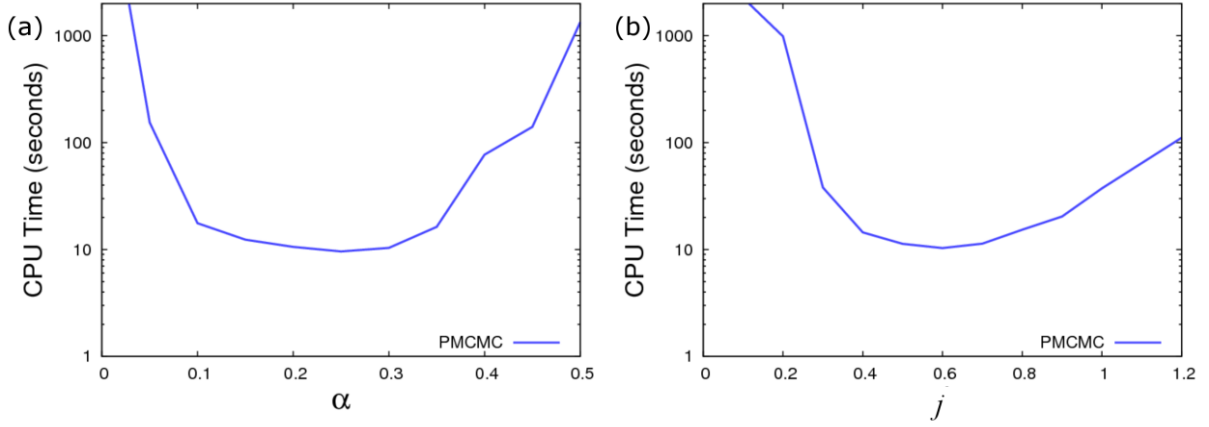

**Figure K:** Optimisation of the PMCMC algorithm. Results are shown for the logistic population model in Section 5.4 (CPU time when 3 population measurements are made). Shows how CPU time for 100 effective samples of $r_b$ varies as a function of (a) the average proposal acceptance rate α (fixing $j$=0.6) and (b) the parameter jumping size $j$ (fixing α=0.25).

### Optimisation

The PMCMC algorithm above has two free parameters which need to be optimised: the jumping size $j$ used in Eq.(K1) and the number of particles $Z$. The former can be fixed to an optimised value and the latter can be tuned to give a certain specified acceptance rate α. This is achieved in the algorithm by introducing a floating point version of the particle number $Z_f$ (such that $Z$ is the integer rounded value of $Z_f$) which is updated in the following manner:

$$Z_f^{new} = Z_f \times 1.02 \qquad \text{if PMCMC proposal accepted,}$$
$$Z_f^{new} = Z_f \times 1.02^{-\frac{\alpha}{1-\alpha}} \quad \text{if PMCMC proposal rejected.} \qquad (K7)$$

(note, this is analogous to the approach used in Eqs.(C2) and (C3)).

Figure K shows the algorithm can be optimised by scanning $j$ and α. In this particular example CPU is minimised when α≈0.25 and $j$≈0.6, with performance not very sensitive to these precise values.

### Appendix L: Effective sample number

Given $X$ correlated MCMC samples of some quantity $x^i$, the autocorrelation function can be approximated by

$$F_\tau = \frac{1}{(X-\tau)\overline{\sigma_x^2}} \sum_{i=1}^{X-\tau} (x^i - \overline{x})(x^{i+\tau} - \overline{x}), \qquad (L1)$$

where estimates for the average and variance of $x$ are given by

$$\overline{x} = \tfrac{1}{X}\sum_{i=1}^{X} x^i, \quad \overline{\sigma_x^2} = \frac{1}{X-1}\sum_{i=1}^{X}(x^i - \overline{x})^2. \qquad (L2)$$

The effective sample size is given by the actual sample number $X$, correcting for correlations between successive samples:

$$X_{eff} = \frac{X}{1+2\sum_{\tau=1}^{\infty} F_\tau}. \tag{L3}$$

When actually calculating $X_{eff}$, clearly the sum in Eq.(L3) cannot go to infinity. In fact, $F_\tau$ often exhibits considerable fluctuations for large $\tau$, and these can generate unwanted bias. The simplest way to deal with these is to truncate the sum in Eq.(L3) up to a maximum size $\tau_{max}$, which is defined to be the largest value of $\tau$ for which the following condition holds true (see [45]):

$$F_\tau > 0.05. \tag{L4}$$

## Appendix M: Details for disease prevalence model

In this appendix we provide additional details relevant to the disease prevalence and diagnostic test model in Section 5.1.

### Simulation and prior details

Simulated data was created using $Se_1=Se_2=0.6$ and $Sp_1=Sp_2=0.9$ for $P=1000$ individuals. Different values of individual number $P$ were used to generate Fig. 5(d). The prior distributions for parameters $Se_1$, $Se_2$ and $p_D$ were assumed to be uniform between 0 and 1, and for $Sp_1$ and $Sp_2$ to be uniform between 0.5 and 1 (the reason this isn't from 0 is because otherwise the posterior becomes bimodal).

### Observation model and latent process likelihood

The model is illustrated in Fig. 5(a). The true disease status of individuals is represented by Bernoulli variables $D_e$, where $D_e=1$ (or 0) denotes that individual $e$ is infected (or uninfected) with probability $p_D$ (or $1-p_D$). The test data $y_e^t$ for test type $t$ are 1 (or 0), indicating a positive (or negative) result.

We identify the following model parameters $\theta=\{p_D, Se_1, Sp_1, Se_2, Sp_2\}$ and latent variables $\xi=\{D\}$. The observation model and latent process likelihood are given by

$$\pi(y|\xi,\theta) = \prod_{t=1,2} Se_t^{N_t^{1|1}} (1-Se_t)^{N_t^{0|1}} Sp_t^{N_t^{0|0}} (1-Sp_t)^{N_t^{1|0}},$$
$$\pi(\xi|\theta) = p_D^{N^1} (1-p_D)^{N^0}, \tag{M1}$$

where $N^d$ is the number of individuals with disease status $d$ and $N_t^{r|d}$ is the number of cases in which test $t$ gives result $r$ for individuals with disease status $d$.

### Importance distributions

Step 2 of the PBP algorithm (introduced in the Section 3.4) makes use of IDs. Successive approximations for these IDs were then discussed in Section 4. Here we explicitly present expressions for this particular model.

$ID_0$ is given by the model itself

$$f_{ID_0}(D_e | D_{e'<e}, \theta) = f_{Bern}(D_e | p_D), \tag{M2}$$

where $f_{Bern}$ is the Bernoulli probability distribution.

$ID_1$ takes into account both the model and the observations. From Eq.(16) this is given by

$$f_{ID_1}(D_e \mid D_{e'<e}, \theta, y) = cf_{Bern}(D_e \mid p_D) \times \prod_{t=1,2} \pi(y_e^t \mid D_e), \qquad (M3)$$

where *c* is a normalisation constant. Explicitly incorporating the observation model from Eq.(M1), this becomes

$$f_{ID_1}(D_e \mid D_{e'<e}, \theta, y) = f_{Bern}(D_e \mid \tfrac{p_1}{p_1+p_0}), \qquad (M4)$$

where

$$p_1 = p_D \times \begin{cases} Se_1 & \text{if obs. 1 +ve} \\ 1-Se_1 & \text{if obs. 1 -ve} \end{cases} \times \begin{cases} Se_2 & \text{if obs. 2 +ve} \\ 1-Se_2 & \text{if obs. 2 -ve} \end{cases}$$

$$p_0 = (1-p_D) \times \begin{cases} 1-Sp_1 & \text{if obs. 1 +ve} \\ Sp_1 & \text{if obs. 1 -ve} \end{cases} \times \begin{cases} 1-Sp_2 & \text{if obs. 2 +ve} \\ Sp_2 & \text{if obs. 2 -ve} \end{cases}. \qquad (M5)$$

In this particular example $ID_1$ (unusually) represents the exact importance distribution (*i.e.* it directly samples from the posterior), so no higher order terms need to be considered.

### Proposals

As an illustration of how PBPs are implemented in practice, we explicitly go through step 2 of the PBP algorithm from Section 3.4 (which stochastically modifies $D_e^i$ to generate $D_e^p$).

The ID is given by a Bernoulli distribution with disease probability *z*:

$$f_{ID}(D_e \mid D_{e'<e}, \theta, y) = f_{Bern}(D_e \mid z). \qquad (M6)$$

For $ID_0$, *z* is equal to $p_D$ and for $ID_1$, $z=p_1/(p_1+p_0)$, where the definitions for $p_0$ and $p_1$ are given in Eq.(M5).

Sequentially going through each individual *e*, the values for the initial $z_e^i$ and proposed $z_e^p$ states are calculated. Table 1 shows that for the Bernoulli distribution:

1) For $z_e^p > z_e^i$: if $D_e^i = 1$ we simply set $D_e^p = 1$, otherwise we draw a random number from *0* to *1* and if it is less than $\frac{z_e^p - z_e^i}{1 - z_e^i}$ we set $D_e^p = 1$ else $D_e^p = 0$.

2) For $z_e^p \leq z_e^i$: if $D_e^i = 0$ we simply set $D_e^p = 0$, otherwise we draw a random number from *0* to *1* and if it is less than $1 - \frac{z_e^p}{z_e^i}$ we set $D_e^p = 0$ else $D_e^p = 1$.

### Gibbs samplers

For the disease diagnostic test model it is possible to explicitly sample directly from the posterior when model parameters and latent variables are each considered separately.

**Model parameters:** The following samples are sequentially drawn from beta distributions

$$p_D \sim Beta(N^1+1, N^0+1),$$
$$Se_1 \sim Beta(N_1^{1|1}+1, N_1^{0|1}+1),$$
$$Sp_1 \sim Beta(N_1^{0|0}+1, N_1^{1|0}+1), \quad (M7)$$
$$Se_2 \sim Beta(N_2^{1|1}+1, N_2^{0|1}+1),$$
$$Sp_2 \sim Beta(N_2^{0|0}+1, N_2^{1|0}+1).$$

In the case of $Sp_1$ and $Sp_2$ samples are rejected if less than 0.5 (to respect the prior), but this occurs very infrequently.

**Latent variables:** Each individual $e$ is considered in turn and, using the definitions in Eq.(M5), we set $D_e=1$ with probability $p_1/(p_1+p_0)$ else $D_e=0$.

# Appendix N: Details of the stochastic volatility model

In this appendix we provide additional details relevant to stochastic volatility model in Section 5.2.

## Simulation and prior details

Simulated data was created using $\mu=-10$, $\phi=0.99$, $\nu=12$, $\sigma^2=0.0121$, and for simplicity the initial condition was set to $h_1=\mu$. Different value of correlation parameter $\phi$ were used to generate Fig. 6(d). The prior distributions for all variables were taken to be flat and in the ranges $-\infty$–$\infty$ for $\mu$ and $h_1$, 0.0001–0.9999 for $\phi$, 2–50 for $\nu$ and 0–$\infty$ for $\sigma^2$.

## Observation model and latent process likelihood

We identify the following model parameters $\theta=\{\mu,\phi,\nu,\sigma^2,h_1\}$ and latent variables $\xi=\{h_{e>1}\}$. The DAG structure is illustrated in Fig. 6(a). The observation model and latent process likelihood are given by

$$\pi(y|\xi,\theta) = \prod_{e=1}^{E} \frac{\Gamma(\frac{\nu+1}{2})}{\sqrt{\nu\pi}\Gamma(\frac{\nu}{2})}\left(1+\frac{y_e^2}{\nu e^{h_e}}\right)^{-\frac{\nu+1}{2}} e^{-h_e/2},$$

$$\pi(\xi|\theta) = \prod_{e=2}^{E} \frac{1}{\sqrt{2\pi\sigma^2}} e^{-\frac{(h_e-\mu-\phi(h_{e-1}-\mu))^2}{2\sigma^2}}, \quad (N1)$$

where $\Gamma$ is the gamma function and $E$ the observation period.

## Importance distributions

$ID_0$ is given by the model

$$f_{ID_0}(h_e|h_{e'<e},\theta) = f_{\text{norm}}\left(h_e|\mu_e^{\text{mod}},\sigma^2\right), \quad (N2)$$

where $\mu_e^{\text{mod}} = \mu + \phi(h_{e-1}-\mu)$.

Equation (16) shows the expression for $ID_1$. The first thing to note is that the product of the model $\pi(\xi_e|\xi_{e'<e},\theta)$ and observation probability $\pi(y_e|\xi_e,\theta)$ distributions from Eq.(N1) is not a standard distribution with which PBPs can be used (*i.e.* it is not listed in Table 1). One way around this problem is to first approximate the observation model as a normal distribution. This is achieved by Taylor series expanding $\log(\pi(y_e|\xi_e,\theta))$ about $\mu_e^{\text{mod}}$ up to second order, leading to

$$\pi(y_e | \xi_e, \theta) \cong f_{\text{norm}}\left(h_e | \mu_e^{\text{obs}}, \sigma_e^{\text{obs2}}\right), \tag{N3}$$

where

$$\mu_e^{\text{obs}} = \mu_e^{\text{mod}} + \tfrac{1}{2}\sigma_e^{\text{obs2}} \frac{v - q_e}{q_e + 1}, \quad \sigma_e^{\text{obs2}} = \frac{2(q_e+1)^2}{q_e(v_e+1)}, \quad q_e = \frac{ve^{\mu_e^{\text{mod}}}}{y_e^2}. \tag{N4}$$

The product of the two normal distributions in Eqs.(N2) and (N3) give

$$f_{ID_1}(h_e | h_{e'<e}, \theta, y) = f_{\text{norm}}(h_e | \tfrac{\mu_e^{\text{obs}}\sigma^2 + \mu_e^{\text{mod}}\sigma_e^{\text{obs2}}}{\sigma_e^{\text{obs2}} + \sigma^2}, \tfrac{\sigma_e^{\text{obs2}}\sigma^2}{\sigma_e^{\text{obs2}} + \sigma^2}). \tag{N5}$$

The definition of ID$_2$ from Eq.(17) is given by

$$f_{ID_2}(h_e | h_{e'<e}, \theta, y) \propto \pi(h_e | h_{e-1}, \theta)\pi(y_e | h_e, \theta)\int \pi(h_{e+1} | h_e, \theta)\pi(y_{e+1} | h_{e+1}, \theta)dh_{e+1}. \tag{N6}$$

In other words, the posterior distribution for $h_e$ at time $e$ depends not only on the value of $h_{e-1}$ (i.e. the previous time point), but also on the observations at times $e$ and $e+1$. In the general case this integral is intractable, but again following the Taylor series expansion approximation to the observation likelihood around $\bar{\mu} = \mu_e^{\text{mod}} = \mu + \phi(h_{e-1} - \mu)$, the following set of approximations can be made:

$$\begin{aligned}
\pi(h_e | h_{e-1}, \theta) &= f_{\text{norm}}\left(h_e | \bar{\mu}, \sigma^2\right), \\
\pi(y_e | h_e, \theta) &\cong f_{\text{norm}}\left(h_e | \mu_e^{\text{obs}}, \sigma_e^{\text{obs2}}\right), \\
\pi(h_{e+1} | h_e, \theta) &= f_{\text{norm}}\left(h_{e+1} | \mu + \phi(h_e - \mu), \sigma^2\right), \\
\pi(y_{e+1} | h_{e+1}, \theta) &\cong f_{\text{norm}}\left(h_{e+1} | \mu_{e+1}^{\text{obs}}, \sigma_{e+1}^{\text{obs2}}\right).
\end{aligned} \tag{N7}$$

where

$$\mu_e^{\text{obs}} = \bar{\mu} + \tfrac{1}{2}\sigma_e^{\text{obs2}} \frac{v - q_e}{q_e + 1}, \quad \sigma_e^{\text{obs2}} = \frac{2(q_e+1)^2}{q_e(v_e+1)}, \quad q_e = \frac{ve^{\bar{\mu}}}{y_e^2},$$

$$\mu_{e+1}^{\text{obs}} = \bar{\mu} + \tfrac{1}{2}\sigma_{e+1}^{\text{obs2}} \frac{v - q_{e+1}}{q_{e+1} + 1}, \quad \sigma_{e+1}^{\text{obs2}} = \frac{2(q_{e+1}+1)^2}{q_e(v_{e+1}+1)}, \quad q_{e+1} = \frac{ve^{\bar{\mu}}}{y_{e+1}^2}. \tag{N8}$$

Substituting the results from Eq.(N7) into (N6) gives

$$f_{ID_2}(h_e | h_{e'<e}, \theta, y) \propto f_{\text{norm}}\left(h_e | \bar{\mu}, \sigma^2\right) f_{\text{norm}}\left(h_e | \mu_e^{\text{obs}}, \sigma_e^{\text{obs2}}\right) \times$$
$$\int f_{\text{norm}}\left(h_{e+1} | \mu + \phi(h_e - \mu), \sigma^2\right) f_{\text{norm}}\left(h_{e+1} | \mu_{e+1}^{\text{obs}}, \sigma_{e+1}^{\text{obs2}}\right) dh_{e+1}. \tag{N9}$$

Integrating over two normally distributed quantities is Gaussian distributed with respect to the difference in means with variance given by the sum of the variances of the two original distributions. Consequently, Eq.(N9) becomes

$$f_{ID_2}(h_e \mid h_{e' < e}, \theta, y) \propto f_{\text{norm}}\left(h_e \mid \overline{\mu}, \sigma^2\right) f_{\text{norm}}\left(h_e \mid \mu_e^{\text{obs}}, \sigma_e^{\text{obs}2}\right) \times e^{-\frac{\left(\mu + \phi(h_e - \mu) - \mu_{e+1}^{\text{obs}}\right)^2}{2\left(\sigma^2 + \sigma_{e+1}^{\text{obs}2}\right)}}. \tag{N10}$$

When written in terms of $h_e$ the last term becomes another normal distribution

$$f_{ID_2}(h_e \mid h_{e' < e}, \theta, y) \propto f_{\text{norm}}\left(h_e \mid \overline{\mu}, \sigma^2\right) f_{\text{norm}}\left(h_e \mid \mu_e^{\text{obs}}, \sigma_e^{\text{obs}2}\right) f_{\text{norm}}\left(h_e \mid \mu_e^{\text{next}}, \sigma_e^{\text{next}2}\right), \tag{N11}$$

with mean and variance that capture information from the next observation (*i.e.* at time *e*+1)

$$\mu_e^{\text{next}} = \mu + \frac{\mu_{e+1}^{\text{obs}} - \mu}{\phi}, \quad \sigma_e^{\text{next}2} = \frac{\sigma^2 + \sigma_{e+1}^{\text{obs}2}}{\phi^2}. \tag{N12}$$

The product of the three normal distributions in Eq.(N11) gives the final result

$$f_{ID_2}(h_e \mid h_{e' < e}, \theta, y) = f_{\text{norm}}(h_e \mid \mu_e^{\text{res}}, \sigma_e^{\text{res}2}), \tag{N13}$$

where

$$\frac{1}{\sigma_e^{\text{res}2}} = \frac{1}{\sigma^2} + \frac{1}{\sigma_e^{\text{obs}2}} + \frac{1}{\sigma_e^{\text{next}2}}, \quad \frac{\mu_e^{\text{res}}}{\sigma_e^{\text{res}2}} = \frac{\overline{\mu}}{\sigma^2} + \frac{\mu_e^{\text{obs}}}{\sigma_e^{\text{obs}2}} + \frac{\mu_e^{\text{next}}}{\sigma_e^{\text{next}2}}. \tag{N14}$$

### The standard approach
By multiplying the two expressions in Eq.(N1), the posterior distribution is given by

$$\pi(\mu, \phi, \nu, \sigma^2, \boldsymbol{h} \mid y) \propto \prod_{e=2}^{E} \frac{1}{\sqrt{2\pi\sigma^2}} e^{-\frac{(h_e - \mu - \phi(h_{e-1} - \mu))^2}{2\sigma^2}} \times \prod_{e=1}^{E} \frac{\Gamma(\frac{\nu+1}{2})}{\sqrt{\nu\pi}\Gamma(\frac{\nu}{2})} \left(1 + \frac{y_e^2}{\nu e^{h_e}}\right)^{-\frac{\nu+1}{2}} e^{-h_e/2}. \tag{N15}$$

Rearranging this gives

$$\pi(\mu \mid y, \phi, \nu, \sigma^2, \boldsymbol{h}) \propto e^{-\frac{(1-\phi)^2(E-1)\left(\mu - \frac{1}{(1-\phi)(E-1)}\sum_{e=2}^{E} h_e - \phi h_{e-1}\right)^2}{2\sigma^2}}, \tag{N16}$$

which takes the form of a standard normal distribution

$$\pi(\mu \mid y, \phi, \nu, \sigma^2, \boldsymbol{h}) = f_{\text{norm}}\left(\mu \mid \frac{1}{(1-\phi)(E-1)}\sum_{e=2}^{E} h_e - \phi h_{e-1}, \frac{\sigma^2}{(1-\phi)^2(E-1)}\right). \tag{N17}$$

Consequently, $\mu$ is sampled in the following way:

$$\mu \sim N\left(\frac{1}{(1-\phi)(E-1)}\sum_{e=2}^{E} h_e - \phi h_{e-1}, \frac{\sigma^2}{(1-\phi)^2(E-1)}\right). \tag{N18}$$

Similarly, the correlation parameter $\phi$ is also sampled from a normal distribution

$$\phi \sim N\left(\frac{\sum_{e=2}^{E}(h_e - \mu)(h_{e-1} - \mu)}{\sum_{e=2}^{E}(h_{e-1} - \mu)^2}, \frac{\sigma^2}{\sum_{e=2}^{E}(h_{e-1} - \mu)^2}\right). \tag{N19}$$

The expression in Eq.(N15) can be rearranged to give

$$\pi(\sigma^2 \mid y, \mu, \phi, \nu, \mathbf{h}) \propto \frac{1}{\sigma^{E-1}} e^{-\frac{\sum_{e=2}^{E}(h_e - \mu - \phi(h_{e-1} - \mu))^2}{2\sigma^2}}, \qquad (N20)$$

which is an inverted chi-squared distribution with respect to $\sigma^2$. All samples generated which have zero prior probability are subsequently resampled.

We adopt a simple random walk Metropolis-Hastings scheme (*i.e.* propose a new parameter / variable by adding a normally distributed contribution to its existing value and accepting or rejecting that change) for $\nu$ and $h_e$. In the case of $h_e$ care is taken to only calculate those parts of the latent process likelihood and observation probability that actually change (to optimise the code as far as possible). The jumping sizes of the separate proposals are individually tuned to give acceptance approximately 33% of the time (using the same procedure as for *j* in Appendix C).

## Appendix O: Details for mixed models

In this appendix we provide additional details relevant to the mixed model in Section 5.3.

### Simulation and prior details
In all cases flat uninformative priors were assumed for parameters.

Here we take *y* to represent measurements of heights within a population. Two fixed effects are assumed: $\beta_1$ represents the average height of females and $\beta_2$ gives the average height difference between males and females. As illustrated in Fig. O1, the model assumes a population of size *P* randomly mated over four generations (which leads to a sparse inverse matrix $\mathbf{A}^{-1}$ [27]). Here individuals in the 1st generation are assumed to be unrelated (*i.e.* conditionally independent) and those in the 2nd, 3rd and 4th generations are conditionally dependent on exactly two individuals in the previous generation (*i.e.* their parents). Simulated data was generated using a population size of *P*=1000, two fixed effects $\boldsymbol{\beta}$={1,0.1}, randomly allocated gender (*i.e.* $X_{i2}$=0 or 1 with equal probability along with $X_{i1}$=1) and one additive genetic effect per individual (*i.e.* **Z=I**).

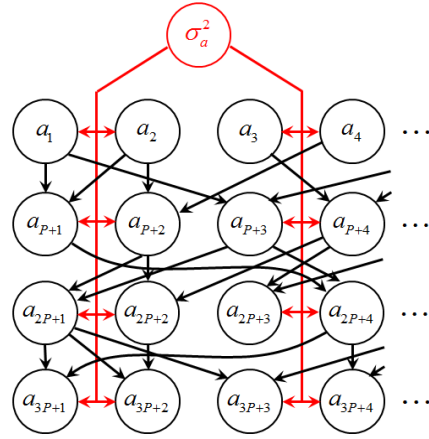

**Figure O1:** A specific quantitative genetics example in which *a* represent additive genetic effects for a population of *P* individuals randomly mated over four generations (note, for clarity β, and *y* have been omitted from this diagram). For the non-founding population the random effect for each individual has contributions from its two parents in the previous generation.

---

[27] Individuals are related to themselves through $A_{nn}$=1 (assuming they are not inbred). If individual *n* is the parent of *p* then $A_{np}$=½, for siblings sharing the same parents $A_{np}$=½, for half-sibs $A_{np}$=¼ and for a grandparent/grandchild relationship $A_{np}$=¼ *etc...* Whilst **A** itself is not sparse, its inverse $\mathbf{A}^{-1}$ is (only diagonal and parent-sibling elements are non-zero).

## Observation model and latent process likelihood

We identify model parameters $\theta = \{\sigma_a^2, \sigma_\varepsilon^2, \boldsymbol{\beta}\}$ and latent variables $\xi = \{\boldsymbol{a}\}$, with residuals $\boldsymbol{\varepsilon}$ incorporated into the observation model.

At first glance it might appear that PBPs are not applicable to this particular model because the latent variables are MVN distributed (*i.e.* a distribution not contained within Table 1). This, however, turns out not to be the case, because of the following transformation.

The latent process likelihood is given by

$$\pi(\xi|\theta) = \pi(\boldsymbol{a}|\sigma_a^2) = \frac{1}{\sqrt{(2\pi\sigma_a^2)^E |A|}} e^{-\frac{1}{2\sigma_a^2}\sum_{d=1}^{E}\sum_{e=1}^{E} a_d A_{de}^{-1} a_e}. \tag{O1}$$

Separating out those terms which depend on $a_E$ in the sum (and remembering that **A** is symmetric and fixed), leads to the product of a normal distribution for $a_E$ (given $a_{e'<E}$) multiplied by a new MVN distribution over the remaining *E-1* latent variables

$$\pi(\boldsymbol{a}|\sigma_a^2) \propto f_{\text{norm}}\left(a_E \,\Big|\, -\sum_{e'=1}^{E-1} A_{Ee'}^{-1} a_{e'},\, \sigma_a^2/A_{EE}^{-1}\right) \times \sigma_a^{-(E-1)} e^{-\frac{1}{2\sigma_a^2}\left(\sum_{d=1}^{E-1}\sum_{e=1}^{E-1} a_d \left(A_{de}^{-1} - A_{Ed}^{-1} A_{Ee}^{-1}/A_{EE}^{-1}\right) a_e\right)}, \tag{O2}$$

where the p.d.f. for the univariate normal distribution is given by

$$f_{\text{norm}}(x|\mu,\sigma^2) = \frac{1}{\sqrt{2\pi\sigma^2}} e^{-\frac{(x-\mu)^2}{2\sigma^2}}. \tag{O3}$$

The scheme above can be iterated until the original MVN distribution is converted into a product of normal distributions for each of the random effects

$$\pi(\boldsymbol{a}|\sigma_a^2) \propto \prod_{e=1}^{E} f_{\text{norm}}(a_e | \mu_e^{\text{mod}}, \sigma_e^{\text{mod}\,2}), \tag{O4}$$

where

$$\mu_e^{\text{mod}} = \sum_{e'<e} M_{ee'} a_{e'}, \quad \sigma_e^{\text{mod}\,2} = s_e \sigma_a^2, \tag{O5}$$

and matrix **M** and vector **s** are fixed and calculated from **A** by recursively applying Eq.(O2). Note, Equation (O4) follows the same structure as Eq.(2), showing that it represents a DAG (specifically, the one illustrated in Fig. 7(a)).

Under the above transformation, the observation model and latent process likelihood are given by

$$\begin{aligned}
\pi(y|\xi,\theta) &= \prod_{i=1}^{N} f_{\text{norm}}\left(y_i \,\Big|\, \sum_{f=1}^{F} X_{if} \beta_f + \sum_{e=1}^{E} Z_{ie} a_e,\, \sigma_\varepsilon^2\right), \\
\pi(\xi|\theta) &= f_{MVN}(\boldsymbol{a}|0, \sigma_a^2 A) \\
&\propto \prod_{e=1}^{E} f_{\text{norm}}(a_e | \mu_e^{\text{mod}}, \sigma_e^{\text{mod}\,2}),
\end{aligned} \tag{O6}$$

where *i* goes over the observations, *f* goes over the fixed effects and *e* goes over the random effects.

## Importance distributions

Different levels of ID approximation are illustrated in Fig. O2.

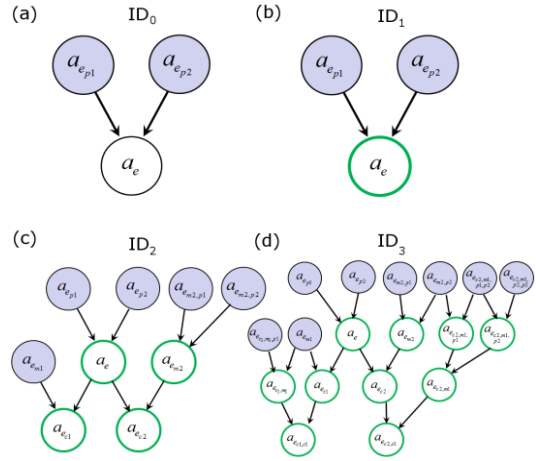

$ID_0$ is given by the model

$$f_{ID_0}(a_e \mid a_{e'<e}, \theta) = f_{norm}\left(a_e \mid \mu_e^{mod}, \sigma_e^{mod2}\right).$$

From Eq.(16), we see that $ID_1$ is generated by taking the product of the model and the observation probability distributions. For simplicity we assume that each observation contains a single random effect, but that each random effect may have many observations made on it (PBPs can also be applied in the more general case, but estimation of the IDs becomes somewhat more complicated).

**Figure O2:** Various levels of approximation used for estimating $f_{ID}(a_e \mid a_{e'<e}, \theta, y)$ for the quantitative genetics model. The shaded circles represent known additive genetic effect $a_{e'<e}$ and the bold, green circles indicate the actual trait measurements used.

A rearrangement of the observation model in Eq.(O4) leads to an effective observation probability for each individual random effect of

$$\pi(y_e \mid a_e, \theta) = f_{norm}(a_e \mid \mu_e^{obs}, \sigma_e^{obs2}), \tag{O7}$$

where $y_e$ combines all observations $y_i$ that include random effect $a_e$, and

$$\mu_e^{obs} = \frac{\sum_i \left(y_i - \sum_f X_{if}\beta_f\right)Z_{ie}}{\sum_i Z_{ie}^2}, \quad \sigma_e^{obs2} = \frac{\sigma_\varepsilon^2}{\sum_i Z_{ie}^2}. \tag{O8}$$

Taking the product of the two normal distributions in Eqs.(O4) and (O7) leads to the final result

$$f_{ID_1}(a_e \mid a_{e'<e}, \theta, y) = f_{norm}(a_e \mid \tfrac{\mu_e^{obs}\sigma_e^{mod2} + \mu_e^{mod}\sigma_e^{obs2}}{\sigma_e^{obs2} + \sigma_e^{mod2}}, \tfrac{\sigma_e^{obs2}\sigma_e^{mod2}}{\sigma_e^{obs2} + \sigma_e^{mod2}}). \tag{O9}$$

The distribution for $ID_2$ takes into account those random effects which depend on $a_e$. As stated in Eq.(17) and illustrated in Fig. O2(c), derivation of $ID_2$ involves integrating over those random effects $a_d$ which depend on $a_e$. Explicitly writing down the expression for $f_{ID_2}(\xi_e \mid \xi_{e'<e}, \theta, y)$ is somewhat verbose. Instead, here we build up the final result by considering different contributions in turn.

To start with we consider a particular random effect $a_d$ (which depends on $a_e$), and find out how it affects $f_{ID_2}$ when it is integrated out. The contribution to the model part of the full posterior from $a_d$ is given by

$$f_{norm}(a_d \mid \sum_{e'} M_{de'}a_{e'}, \sigma_d^{mod2}), \tag{O10}$$

where $e'$ sums over all those random effects on which $a_d$ depends. Three possibility exist for $e'$: 1) $e' < e$, in which case $a_{e'}$ is known (as represented by the shaded circles in Fig. O2), 2) $e'=e$ and 3) $e' > e$, in which case the additional latent variable *e'* has to first be integrated out. In the case of the third option, Eq.(O10) is first recast in terms of the specific variable *e'=r* that needs to be integrated out:

$$f_{norm}(a_r \mid \frac{a_d - \sum_{e' \neq r} M_{de'} a_{e'}}{M_{dr}}, \frac{\sigma_d^{mod2}}{M_{dr}^2}). \tag{O11}$$

Now we remember that the posterior also has a contribution for $a_r$ coming from its measurement and those random effects upon which it depends. These are captured by $ID_1$ from Eq.(O9), which is given by

$$f_{norm}(a_r \mid \mu_r^{post}, \sigma_r^{post2}), \tag{O12}$$

where

$$\mu_r^{post} = \frac{\mu_r^{obs} \sigma_r^{mod2} + \mu_r^{mod} \sigma_r^{obs2}}{\sigma_r^{obs2} + \sigma_r^{mod2}}, \qquad \sigma_r^{post2} = \frac{\sigma_r^{obs2} \sigma_r^{mod2}}{\sigma_r^{obs2} + \sigma_r^{mod2}}. \tag{O13}$$

Multiplying Eqs.(O11) and (O12) and integrating over $a_r$ leads to the posterior probability being proportional to

$$e^{-\frac{1}{2\left(\frac{\sigma_d^{mod2}}{M_{dr}^2} + \sigma_r^{post2}\right)} \left(\frac{a_d - \sum_{e' \neq r} M_{de'} a_{e'}}{M_{dr}} - \mu_r^{post}\right)^2}. \tag{O14}$$

This can again be re-cast in terms of $a_d$:

$$f_{norm}(a_d \mid M_{dr} \mu_r^{post} + \sum_{e' \neq r} M_{de'} a_{e'}, \sigma_d^{mod2} + M_{dr}^2 \sigma_r^{post2}). \tag{O15}$$

Compared to the original expression in Eq.(O10), we see that the effect of integrating out $a_r$ is to replace $a_r$ with the posterior estimate $\mu_r^{post}$ in the mean and to add an additional contribution to the variance. The procedure above can be repeated for all *e'>e*, leading to

$$f_{norm}(a_d \mid \sum_{e'<e} M_{de'} a_{e'} + M_{ee} a_e + \sum_{e'>e} M_{de'} \mu_{e'}^{post}, \sigma_d^{mod2} + \sum_{e'>e} M_{de'}^2 \sigma_{e'}^{post2}). \tag{O16}$$

We now introduce the contribution which comes from the observation on $a_d$ itself:

$$f_{norm}(a_d \mid \mu_d^{obs}, \sigma_d^{obs2}). \tag{O17}$$

Multiplying Eqs.(O16) and (O17) and integrating over $a_d$ implies that the posterior is proportional to

$$e^{-\frac{1}{2\left(\sigma_d^{mod2} + \sigma_d^{obs2} + \sum_{e'>e} M_{de'}^2 \sigma_{e'}^{post2}\right)} \left(\sum_{e'<e} M_{de'} a_{e'} + M_{ee} a_e + \sum_{e'>e} M_{de'} \mu_{e'}^{post} - \mu_d^{obs}\right)^2}. \tag{O18}$$

This can be recast in terms of $a_e$:

$$f_{norm}(a_e \mid \mu_{d \to e}, \sigma_{d \to e}^2), \tag{O19}$$

where

$$\mu_{d \to e} = \frac{\mu_d^{\text{obs}} - \left( \sum_{e'<e} M_{de'} a_{e'} + \sum_{e'>e} M_{de'} \mu_{e'}^{\text{post}} \right)}{M_{ee}},$$

$$\sigma_{d \to e}^2 = \frac{\sigma_d^{\text{mod}2} + \sigma_d^{\text{obs}2} + \sum_{e'>e} M_{de'}^2 \sigma_{e'}^{\text{post}2}}{M_{ee}^2}.$$

(O20)

Equation (O19) represents the overall contribution to $f_{ID2}$ from latent variable $d$. To find $f_{ID2}$, therefore, this distribution must be multiplied over all random effects $d$ which depend on $a_e$, and also the observation and model contributions from $a_e$ itself must be included:

$$f_{\text{norm}}(a_e \mid \mu_e^{\text{post}}, \sigma_e^{\text{post}2}) \prod_d f_{\text{norm}}(a_e \mid \mu_{d \to e}, \sigma_{d \to e}^2). \tag{O21}$$

Multiplication of these normal distributions yield the final result

$$f_{ID_2}(a_e \mid a_{e'<e}, \theta, y) = f_{\text{norm}}(a_e \mid \mu_e^{ID}, \sigma_e^{ID2}), \tag{O22}$$

where

$$\mu_e^{ID} = \left( \frac{\mu_e^{\text{post}}}{\sigma_e^{\text{post}2}} + \sum_d \frac{\mu_{d \to e}}{\sigma_{d \to e}^2} \right) \sigma_e^{ID2},$$

$$\frac{1}{\sigma_e^{ID2}} = \frac{1}{\sigma_e^{\text{post}2}} + \sum_d \frac{1}{\sigma_{d \to e}^2}.$$

(O23)

### Gibbs samplers

Mixed models represent a case for which it is possible to explicitly sample directly from the posterior when model parameters and latent variables are each considered separately [37]. Assuming a simple uniform prior[28], multiplying the two expressions in Eq.(O6) leads to the posterior probability distribution

$$\pi(\sigma_a^2, \sigma_\varepsilon^2, \boldsymbol{\beta}, \boldsymbol{a} \mid \boldsymbol{y}) \propto \frac{1}{\sqrt{(2\pi\sigma_a^2)^E |A|}} e^{-\frac{1}{2\sigma_a^2} \boldsymbol{a}^T A^{-1} \boldsymbol{a}} \prod_{i=1}^N \frac{1}{\sqrt{2\pi\sigma_\varepsilon^2}} e^{-\frac{1}{2\sigma_\varepsilon^2} \left( y_i - \sum_f X_{if} \beta_f - \sum_e Z_{ie} a_e \right)^2}. \tag{O24}$$

The following Gibbs proposals can be identified, which are sequentially applied to constitute a single "update":

**Model parameters:** Rearranging (O24) gives

$$\pi(\sigma_a^2 \mid \boldsymbol{y}, \sigma_\varepsilon^2, \boldsymbol{\beta}, \boldsymbol{a}) \propto \sigma_a^{-E} e^{-\frac{1}{2\sigma_a^2} \boldsymbol{a}^T A^{-1} \boldsymbol{a}}. \tag{O25}$$

That is, with all other quantities fixed the posterior has an inverted chi-squared distribution with respect to $\sigma_a^2$. Similarly, we find that

$$\pi(\sigma_\varepsilon^2 \mid \boldsymbol{y}, \sigma_a^2, \boldsymbol{\beta}, \boldsymbol{a}) \propto \sigma_\varepsilon^{-N} e^{-\frac{1}{2\sigma_\varepsilon^2} \sum_{i=1}^{N}\left(y_i - \sum_f X_{if}\beta_f - \sum_e Z_{ie}a_e\right)^2}, \tag{O26}$$

and so $\sigma_\varepsilon^2$ also has an inverted chi-squared distribution. See below for how to draw samples from the inverted chi-squared distributions in Eqs.(O25) and (O26).

---

[28] Non-uniform priors can be easily be implemented provided they also take an inverted chi-squared distribution.

Taking each fixed effect $\beta_f$ in turn, the posterior can be written as

$$\pi(\beta_f \mid \boldsymbol{y}, \sigma_a^2, \sigma_\varepsilon^2, \beta_{f'\neq f}, \boldsymbol{a}) = f_{\text{norm}}\left(\beta_f \mid \sum_i X_{if}\left[y_i - \sum_{f'\neq f} X_{if'}\beta_{f'} - \sum_e Z_{ie}a_e\right], \frac{\sigma_\varepsilon^2}{\sum_i X_{if}^2}\right). \tag{O27}$$

Thus, its value can be sampled from the following normal distribution

$$\beta_f \sim N\left(\sum_i X_{if}\left[y_i - \sum_{f'\neq f} X_{if'}\beta_{f'} - \sum_e Z_{ie}a_e\right], \frac{\sigma_\varepsilon^2}{\sum_i X_{if}^2}\right). \tag{O28}$$

**Latent variables:** Gibbs sampling for random effect $a_e$ is achieved through

$$a_e \sim N\left(\mu_e^{\text{gibbs}}, \sigma_e^{\text{gibbs2}}\right), \tag{O29}$$

where

$$\mu_e^{\text{gibbs}} = \left(\frac{1}{\sigma_\varepsilon^2}\sum_i Z_{ie}\left[y_i - \sum_f X_{if}\beta_f - \sum_{e'\neq e} Z_{ie}a_{e'}\right] - \frac{1}{\sigma_a^2}\sum_{e'\neq e} A_{ee'}^{-1}a_{e'}\right),$$
$$\frac{1}{\sigma_e^{\text{gibbs2}}} = \frac{1}{\sigma_a^2} + \frac{\sum_i Z_{ie}^2}{\sigma_\varepsilon^2}. \tag{O30}$$

**Sampling from the inverse chi-squared distribution**

We assume an inverse chi-squared distribution of the form

$$f_\chi(x \mid N, S) \propto x^{-\frac{M}{2}} e^{-\frac{S}{2x}}. \tag{O31}$$

A simple method to calculate samples from this distribution is through

$$x = \frac{S}{2\rho}, \quad \rho = -\sum_{m=1}^{M-1} \log(u_m), \tag{O32}$$

where $u_m$ are uniform randomly generated numbers between 0 and 1.

# Appendix P: Details for the logistic population model

In this appendix we provide additional details relevant to the logistic population model in Section 5.4.

### Simulation and prior details

Simulated data was generated using the following parameters: birth rate $r_b$=0.6, mortality $\mu$=0.3, carrying capacity $K$=100, and capture probability $p$=0.5. The following priors were used: A gamma distributed prior on $\mu$ with mean 0.3 and variance 0.0144, a beta distributed prior on $p$ with mean 0.5 and variance 0.0025, a uniform prior on $K$ between 0 and 200, and a uniform prior on $r_b$ between 0 and 2.

### Observation model and latent process likelihood

We identify model parameters $\theta = \{r_b, \mu, K, p\}$ and latent variables $\xi = \{b_t, d_t\}$, which give the number of births and deaths during each time interval $t$.

The observation model and latent process likelihood are given by

$$\pi(y \mid \xi, \theta) = \prod_{m=1}^{M} \frac{P_{m_t}!}{y_m!(P_{m_t} - y_m)!} p^{y_m} (1-p)^{P_{m_t} - y_m},$$

$$\pi(\xi \mid \theta) \propto \prod_{t=1}^{T} \frac{\lambda_t^{b_t}}{b_t!} e^{-\lambda_t} \times \frac{v_t^{d_t}}{d_t!} e^{-v_t}, \tag{P1}$$

where $m$ goes over all the measurements, $y_m$ are the number of animals observed at time $m_t$, $P_t$ is the population size, and $\lambda_t = \tau r_b P_t (1 - P_t/K)$ and $v_t = \tau \mu P_t$ are the expected number of births and deaths in time interval $t$.

### The standard approach

Random walk MH updates are used for the parameters $r_b$, $\mu$, $K$, and $p$ (*i.e.* this consists of proposing a new parameter by adding a normally distributed contribution to its existing value and accepting or rejecting that change). The jumping sizes of these separate proposals are individually tuned to give acceptance approximately 33% of the time (using the same procedure as for $j$ in Appendix C).

Regarding the latent variables, four types of proposal are used: 1) incrementing or decrementing a birth number $b_t$ with randomly selected time $t$, 2) doing the same for a randomly selected death number $d_t$, and 3) scanning from $t$=1 to $t$=$T$ and locally incrementing or decrementing both birth number $b_t$ and death number $d_t$ (leaving population sizes unchanged), and 4) scanning from $t$=2 to $t$=$T$ and locally incrementing or decrementing the population size $P_t$ (with corresponding adjustments to $b_t,d_t$ and $b_{t-1},d_{t-1}$). Note these last two options are scanned across all times because here individual proposals are fast (these local changes do not require the entire likelihood and observation model to be calculated).